\documentclass[twocolumn,tighten,twocolumn]{aastex631}
\usepackage{wrapfig}
\usepackage{graphics,graphicx}
\usepackage{epstopdf}
\usepackage[utf8]{inputenc}
\usepackage{tikz}
\usetikzlibrary{shapes,arrows}
\usepackage{amssymb, amsmath, amsthm}
\usepackage[normalem]{ulem}
\usepackage{float}
\usepackage{color}
\usepackage{xcolor}
\usepackage{amsmath} 
\usepackage{graphicx}
\usepackage{booktabs}
\usepackage{multirow}
\usepackage{soul}
\definecolor{ultramarine}{rgb}{0.01, 0.64, 0.86} 
\def\green#1 {{\textcolor{ultramarine}{#1}}\ }

\shorttitle{Dust and GMCs in M31}

\begin{document}

\correspondingauthor{Sihan Jiao, Jingwen Wu, Hauyu Baobab Liu, Chao-Wei Tsai}
\email{sihanjiao@nao.cas.cn, jingwen@nao.cas.cn, baobabyoo@gmail.com, cwtsai@nao.cas.cn}

\title{Deep Andromeda JCMT-SCUBA2 Observations. The Submillimeter Maps and Giant Molecular Clouds}

\author[0000-0002-9151-1388]{Sihan Jiao}
\affil{
National Astronomical Observatories, Chinese Academy of Sciences, 20A Datun Road, Chaoyang District, Beijing 100012, China
}
\affil{
Max Planck Institute for Astronomy, Konigstuhl 17, D-69117 Heidelberg, Germany
}

\author{Jingwen Wu}
\affil{University of Chinese Academy of Sciences, Beijing 100049, China}
\affil{
National Astronomical Observatories, Chinese Academy of Sciences, 20A Datun Road, Chaoyang District, Beijing 100012, China
}

\author[0000-0003-2300-2626]{Hauyu Baobab Liu}
\affiliation{Department of Physics, National Sun Yat-Sen University, No. 70, Lien-Hai Road, Kaohsiung City 80424, Taiwan, R.O.C.}

\author[0000-0002-9390-9672]{Chao-Wei Tsai} 
\affiliation{
National Astronomical Observatories, Chinese Academy of Sciences, 20A Datun Road, Chaoyang District, Beijing 100012, China
} 
\affiliation{Institute for Frontiers in Astronomy and Astrophysics, Beijing Normal University, Beijing 102206, China} 
\affiliation{School of Astronomy and Space Science, University of Chinese Academy of Sciences, Beijing 100049, 
China}

\author{Yuxin Lin}
\affil{Max-Planck-Institut f\"ur Extraterrestrische Physik, Giessenbachstr. 1, D-85748 Garching bei M\"unchen, Germany}

\author{Di Li}
\affil{New Cornerstone Science Laboratory, Department of Astronomy, Tsinghua University, Beijing 100084, China}
\affil{
National Astronomical Observatories, Chinese Academy of Sciences, 20A Datun Road, Chaoyang District, Beijing 100012, China
}

\author{Zhi-Yu Zhang}
\affiliation{School of Astronomy and Space Science, Nanjing University, Nanjing 210093, China}
\affiliation{Key Laboratory of Modern Astronomy and Astrophysics, Ministry of Education, Nanjing 210093, China}

\author[0000-0002-0786-7307]{Yu Cheng}
\affil{National Astronomical Observatory of Japan, 2-21-1 Osawa, Mitaka, Tokyo 181-8588, Japan}

\author{Linjing Feng}
\affil{
National Astronomical Observatories, Chinese Academy of Sciences, 20A Datun Road, Chaoyang District, Beijing 100012, China
}
\affil{University of Chinese Academy of Sciences, Beijing 100049, China}

\author{Henrik Beuther}
\affil{
Max Planck Institute for Astronomy, Konigstuhl 17, D-69117 Heidelberg, Germany
}

\author{Junzhi Wang}
\affil{
School of Physical Science and Technology, Guangxi University, Nanning 530004, China
}

\author{Lihwai Lin}
\affil{
1 Institute of Astronomy \& Astrophysics, Academia Sinica, Taipei 10617, Taiwan
}

\author[0000-0002-8760-6157]{Jakob den Brok}
\affil{Center for Astrophysics $\mid$ Harvard \& Smithsonian, 60 Garden Street, Cambridge, 02138, USA}

\author{Ludan Zhang}
\affil{
National Astronomical Observatories, Chinese Academy of Sciences, 20A Datun Road, Chaoyang District, Beijing 100012, China
}
\affil{University of Chinese Academy of Sciences, Beijing 100049, China}

\author[0000-0001-5950-1932]{Fengwei Xu}
\affil{Kavli Institute for Astronomy and Astrophysics, Peking University, Beijing 100871, People's Republic of China}
\affil{Department of Astronomy, School of Physics, Peking University, Beijing, 100871, People's Republic of China}

\author{Fanyi Meng}
\affil{Department of Astronomy, Tsinghua University, Beijing 100084, People’s Republic of China}
\affil{University of Chinese Academy of Sciences, Beijing 100049, China}

\author{Zongnan Li}
\affil{National Astronomical Observatory of Japan, 2-21-1 Osawa, Mitaka, Tokyo, 181-8588, Japan}
\affil{East Asian Core Observatories Association (EACOA) Fellow}

\author{Ryan P. Keenan}
\affil{
Max Planck Institute for Astronomy, Konigstuhl 17, D-69117 Heidelberg, Germany
}

\author[0000-0002-3462-4175]{Si-Yue Yu}
\affil{Kavli Institute for the Physics and Mathematics of the Universe (Kavli IPMU, WPI), UTIAS, Tokyo Institutes for Advanced Study,
University of Tokyo, Chiba 277-8583, Japan}
\affil{Department of Astronomy, School of Science, The University of Tokyo, 7-3-1 Hongo, Bunkyo, Tokyo 113-0033, Japan}

\author{Niankun Yu}
\affil{Max-Planck-Institut f$\ddot{u}$r Radioastronomie, Auf dem H$\ddot{u}$gel 69, 53121 Bonn, Germany}
\affil{
National Astronomical Observatories, Chinese Academy of Sciences, 20A Datun Road, Chaoyang District, Beijing 100012, China
}

\author[0009-0005-9546-4573]{Zheng Zheng}
\affil{
National Astronomical Observatories, Chinese Academy of Sciences, 20A Datun Road, Chaoyang District, Beijing 100012, China
}
\affil{Key Laboratory of Radio Astronomy and Technology, Chinese Academy of Sciences, Beijing, 100101, China}

\author{Junhao Liu}
\affil{National Astronomical Observatory of Japan, 2-21-1 Osawa, Mitaka, Tokyo, 181-8588, Japan}

\author{Yuxiang Liu}
\affil{University of Chinese Academy of Sciences, Beijing 100049, China}
\affil{
National Astronomical Observatories, Chinese Academy of Sciences, 20A Datun Road, Chaoyang District, Beijing 100012, China
}

\author{Hao Ruan}
\affil{University of Chinese Academy of Sciences, Beijing 100049, China}
\affil{
National Astronomical Observatories, Chinese Academy of Sciences, 20A Datun Road, Chaoyang District, Beijing 100012, China
}

\author{Fangyuan Deng}
\affil{University of Chinese Academy of Sciences, Beijing 100049, China}
\affil{
National Astronomical Observatories, Chinese Academy of Sciences, 20A Datun Road, Chaoyang District, Beijing 100012, China
}

\author{Yuanzhen Xiong}
\affil{University of Chinese Academy of Sciences, Beijing 100049, China}
\affil{
National Astronomical Observatories, Chinese Academy of Sciences, 20A Datun Road, Chaoyang District, Beijing 100012, China
}

\shortauthors{Jiao et al.}

\begin{abstract}
We have carried out unprecedentedly deep, nearly confusion-limited JCMT-SCUBA2 mapping observations on the nearest spiral galaxy, M31 (Andromeda).
The 850\,$\mu$m image with a $\sim$50 pc resolution yields a comprehensive catalog of 383 giant molecular clouds (GMCs) that are associated with the spiral arms.
In addition, it unveiled a population of 189 compact inter-arm GMCs in M31, which are mostly unresolved or marginally resolved.
The masses of all these GMCs are in the range of 2$\times$10$^4$ -- 6$\times$10$^6$ $M_{\odot}$; the sizes are in the range of 30--130 pc.
They follow a mass-size correlation, $M$ $\propto$ $R_{c}$$^{2.5}$.
The inter-arm GMCs are systematically less massive, more diffuse, colder, and have lower star-forming efficiency (SFE) than on-arm GMCs.
Moreover, within individual spatially resolved on-arm and off-arm M31 GMCs, the SFE is considerably lower than the SFE in molecular clouds in main sequence and green valley galaxies.
Follow-up investigations on M31 GMCs may provide clues for how star formation may be quenched in galactic environments. 
Finally, we reconstrained the dust opacity spectral index $\beta$ in the M31 galaxy by combining our new JCMT observations with archival {\it Herschel} and {\it Planck} data and found that the radial variation of $\beta$ may not be as large as was proposed by previous studies.  
\end{abstract}


\section{Introduction}

\begin{figure*}[ht!]
\vspace{-0.3cm}
\hspace{0.5cm}
\includegraphics[width=17cm,]{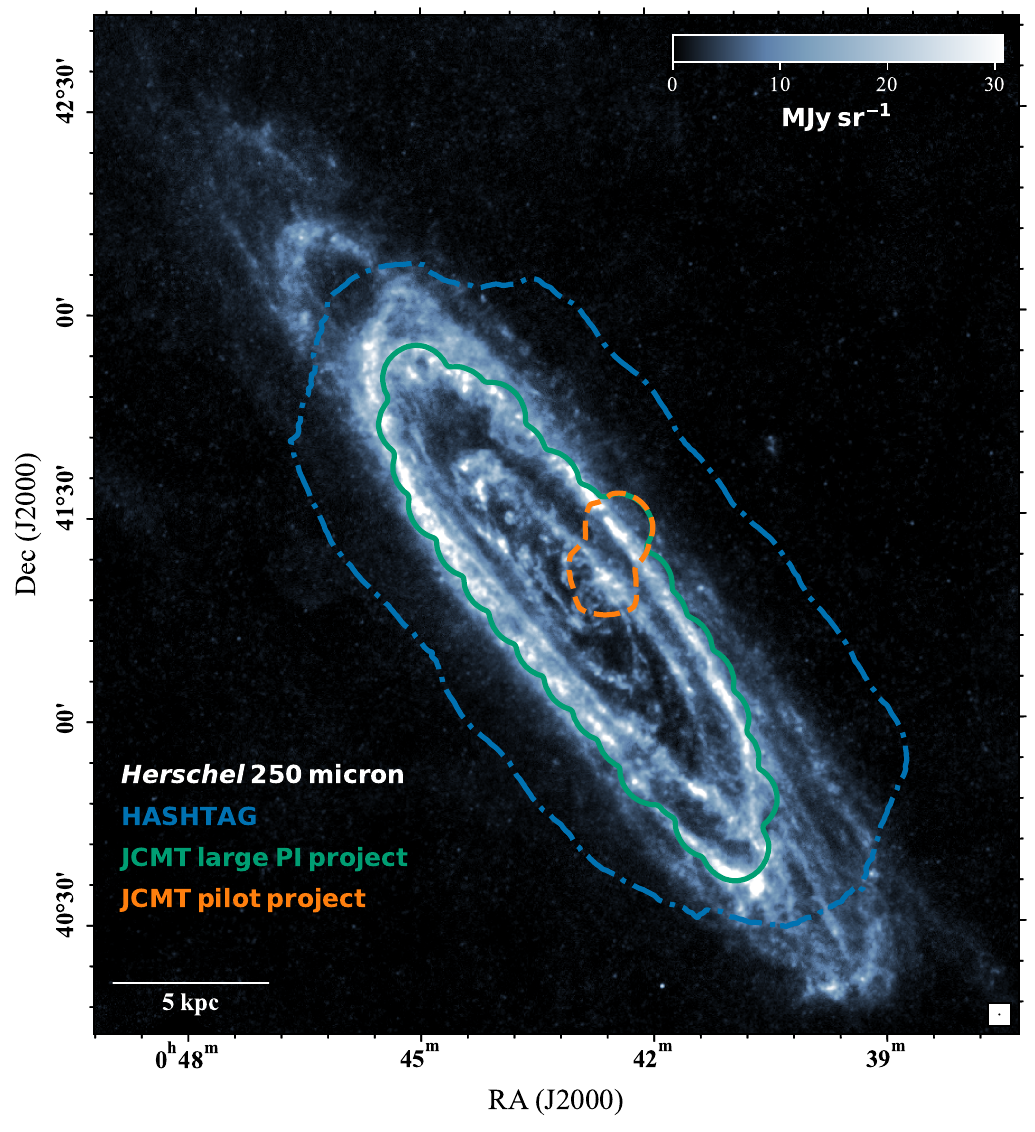}
\caption{
The background color map shows the \textit{Herschel} 250 $\mu$m image \citep{Fritz2012A&A,Smith2012ApJ}. The field of view of the HASHTAG survey \citep{SmithApJS} is outlined in blue dot–dashed lines, while those of the JCMT large PI project and JCMT pilot project are shown in green solid and orange dashed lines, respectively.
}
\label{fig:m31_obs}
\end{figure*}

Giant molecular clouds (GMCs) are the major birthplace for stars \citep{McKee2007ARA&A}.
They also represent fundamental structural units within galaxies.
The study of GMC property variation is essential for understanding the environments in which star formation occurs and how it affects the galaxy's evolution.

\begin{figure*}
\vspace{-0.1cm}
\begin{center}
\includegraphics[width=17cm]{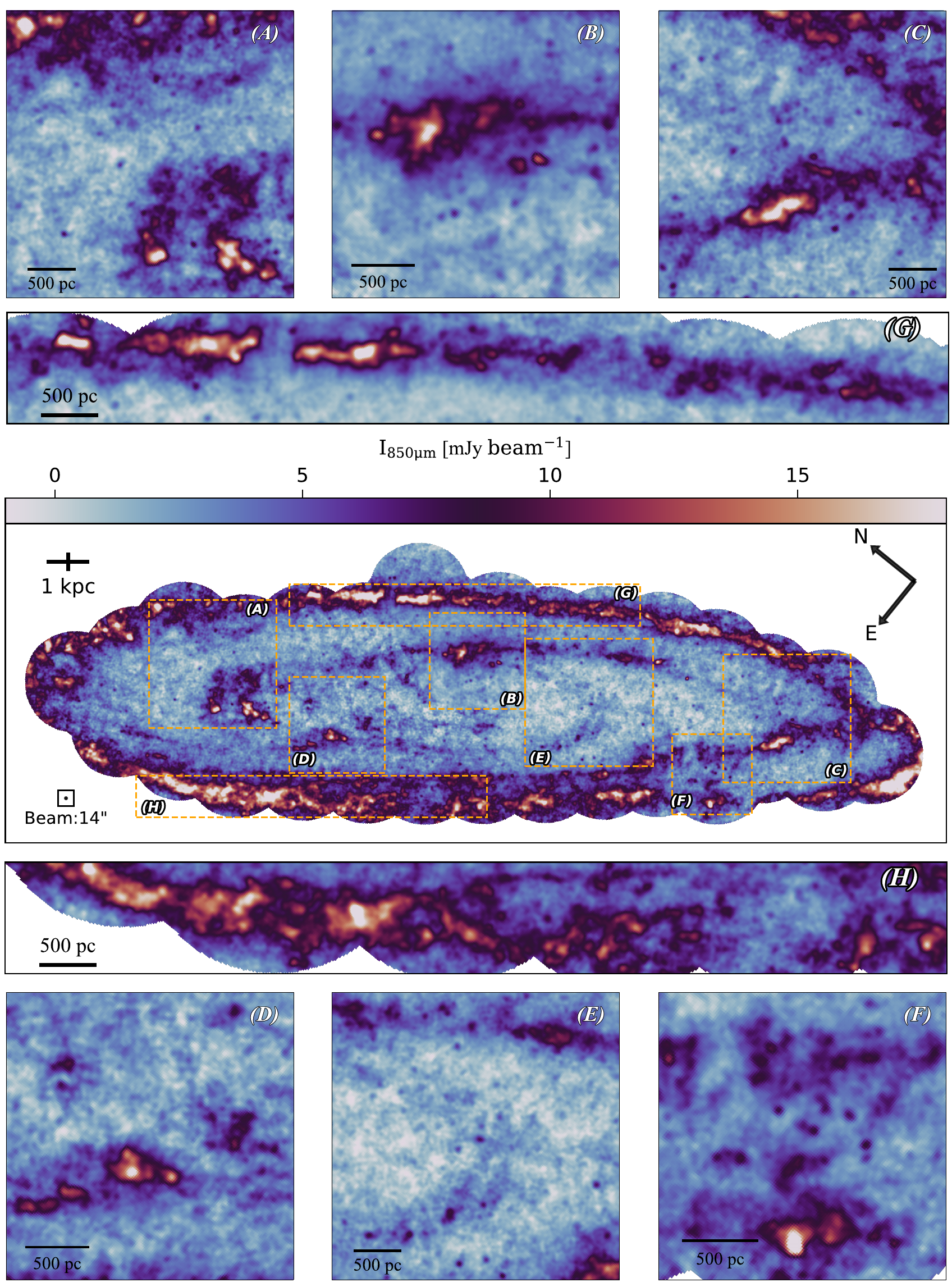}
\end{center}
\caption{
The central panel shows the 850 $\mu$m continuum image of M31 at 14$''$ resolution, produced by combining {\it Planck} and JCMT-SCUBA2 observations.
Orange dashed rectangles highlight eight zoomed-in, selected regions (A-H), to reveal detailed substructures.
Regions A-F focus on inter-arm areas, while G and H display features within the spiral arm.
}
\label{fig:scuba2_850}
\end{figure*}
Our understanding of star formation has relied on detailed studies of GMCs in the Milky Way, using both dust continuum measurements \citep[e.g.,][]{Planck2014A&A,Chen2020MNRAS} and extensive CO emission surveys \citep[e.g.,][]{Dame2001ApJ,Jackson2006ApJS,Colombo2019MNRAS}.
However, the connection between GMCs and large-scale galactic structures, such as spiral arms, remains poorly understood. 
The edge-on projection of the Milky Way and the ambiguity of line-of-sight distances pose significant challenges to achieving a comprehensive, panoramic view of GMCs, particularly in distinguishing GMCs located in inter-arm regions. 

Extragalactic studies provide an opportunity to overcome these challenges, avoiding the foreground/background confusion and distance ambiguities inherent to Galactic observations.
However, except for the galaxies in the Local Group (${\lesssim}3\,$Mpc), a high spatial resolution ($\sim$ 50 pc) to resolve individual GMCs at extragalactic distances requires interferometer observations, which are dominated by observations of low-$J$ rotational CO line emission.
For example, \cite{Rosolowsky2007ApJ} and \cite{Sheth2008ApJ} presented Berkeley Illinois Maryland Association (BIMA) millimeter interferometer observations of GMCs along the spiral arm in M31, while part of this galaxy has been surveyed by CARMA \citep{Caldu-Primo2016AJ}.
\cite{Colombo2014ApJ} used the Plateau de Bure Interferometer (PdBI) to study GMCs across the arms, inter-arm regions, and central disk of M51 (the PdBI Arcsecond Whirlpool Survey), revealing significant environmental variations in cloud populations.
The advent of the Atacama Large Millimeter/submillimeter Array (ALMA) has revolutionized extragalactic studies, enabling surveys of CO emission at cloud-scale resolution across entire galactic disks within hours \citep{Rosolowsky2021MNRAS}.
The PHANGS-ALMA survey \citep{Leroy2021ApJS}, the largest, uniform study of extragalactic gas in nearby galaxies, observed 90 star-forming galaxies in $^{12}$CO (2-1) emission at 50-150 pc spatial resolution.

While CO observations dominate extragalactic GMC studies, dust continuum measurements offer several advantages.
Dust is well mixed with molecular hydrogen (though subject to uncertainties in the dust-to-gas ratio), accounts for a larger proportion of mass in GMCs \citep[e.g.,][]{Draine2011,Planck2011A&A...536A..19P}, and dust emissions are not subject to excitation conditions or chemical abundances.
Furthermore, multi-band dust observations provide dust temperature information for GMCs. 

Despite these advantages, extragalactic dust measurements in GMC-scale-resolution studies remain scarce. 
A notable exception is the study by \cite{Forbrich2020ApJ}, focusing on spiral arm regions identified from giant molecular associations (GMAs) \citep{Kirk2015SSRv} in M31.
By performing Submillimeter Array (SMA) 230 GHz observations, they obtained the continuum detections of the dust emission for 23 GMCs.
These observations were part of an SMA Large Program aimed at conducting a simultaneous survey of 230 GHz continuum and CO line emission from individual GMCs in M31.
The physical properties of M31 GMCs have also been studied in detail through SMA CO observations \citep{Sebastien2021ApJ,Lada2024ApJ,Lada2025ApJ}.
In addition, the JCMT Large Program HASHTAG\footnote{https://hashtag.astro.cf.ac.uk/DR1.html} \citep[HARP and SCUBA-2 High Resolution Terahertz Andromeda Galaxy Survey;][]{Li2020MNRAS,SmithApJS,Deng2025MNRAS}mapped the entire M31 galaxy and primarily detected dust emission in the spiral-arm regions.

To establish a systematic GMC catalog within M31, including also the inter-arm region, we conducted a JCMT-SCUBA2 survey covering the disk out to a radius of $\sim$15 kpc. 
With spatial resolutions of $\sim$30\,pc at 450\,$\mu$m and $\sim$50\,pc at 850\,$\mu$m, our survey matches the typical size of GMCs in the Milky Way \citep{Solomon1979ApJ}.
This dataset enables one of the most extensive dust-based GMC censuses to date in a spiral galaxy, providing uniform coverage of both spiral-arm and inter-arm regions.
By comparing the properties of these GMC populations, we aim to shed light on the role of galactic environments in shaping GMC properties and star formation processes in M31.

This paper is organized as follows. 
Section \ref{section:obs} outlines the observations and the methodologies employed to analyze the data.
Section \ref{section:dust_map} presents a detailed analysis of the dust and gas properties.
The GMC catalog is introduced in Section \ref{sec:gmcs}.
Finally, our main findings are summarized in Section \ref{section:summary}.

\section{OBSERVATIONS and methods}\label{section:obs}

\subsection{JCMT-SCUBA2 Observations}\label{section:obs_jcmt}

Prior to the JCMT large PI project, we conducted pilot SCUBA-2 observations of M31 (JCMT project code M17AP007 and M17BP040, PI: Sihan Jiao), which were carried out over the period of Feb.-Oct. of 2017. 
We utilized two constant velocity (CV) Daisy scanning fields to cover from the center to $\sim$15 kpc radius, achieving an RMS noise of $\sim$1 mJy\,beam$^{-1}$ at 850 $\mu$m, which is close to the confusion limit (0.8 mJy\,beam$^{-1}$) determined by \cite{Geach2017MNRAS}.
The field of view of the pilot observations is outlined in orange dashed lines in Figure \ref{fig:m31_obs}.
Based on this extraordinary deep 850 $\mu$m continuum, we discovered a large number of compact ($<$ 54 pc at the distance of M31) molecular clouds in the on-arm and inter-arm regions. 
To characterize the inter-arm clouds in detail, we performed follow-up observations by using the IRAM 30 m telescope (Project ID: 059-18, PI: Sihan Jiao), NOrthern Extended Millimeter Array (NOEMA, Project ID: W18BW, PI: Sihan Jiao), and ALMA (Project code: 2019.1.01847.S, PI: Sihan Jiao).
A comprehensive analysis of the inter-arm clouds in this pilot region will be presented in Paper II (Jiao et al. in prep.).

\begin{figure*}[ht!]
\vspace{-0.3cm}
\hspace{0.5cm}
\includegraphics[width=17cm,]{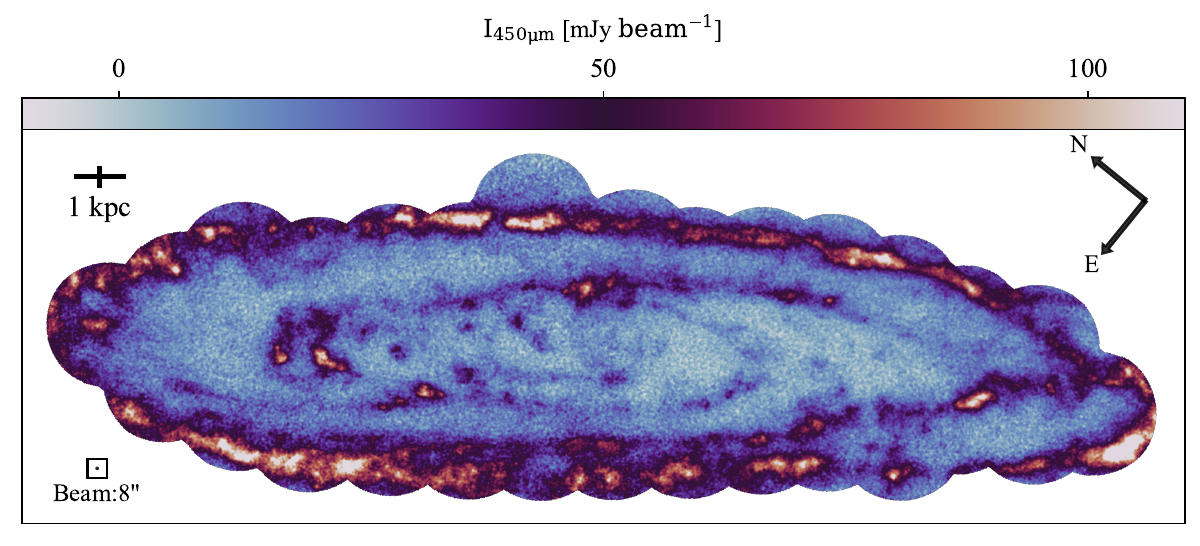}
\caption{
The 8$''$ resolution, 450 $\mu$m continuum image of M31 produced by combining \textit{Herschel} and JCMT-SCUBA2 observations.
}
\label{fig:scuba2_450}
\end{figure*}

Aiming to establish a comprehensive and detailed survey toward inter-arm GMCs in M31, we complete a JCMT large PI project (JCMT project code S19BP002, PI: Jingwen Wu) to survey the M31 inter-arm regions with $\sim$280 hours SCUBA2 observations reaching a similar sensitivity, $<$1.5 mJy\,beam$^{-1}$, at 850 $\mu$m band as in the previous pilot study. 
The observed region covers an area of $\sim$ 1.7$^{\circ}$ $\times$ 0.6$^{\circ}$, from galactic center to galactocentric radius R $\sim$ 15 kpc, with SCUBA2 \citep{Holland2013MNRAS} at 450 and 850 $\mu$m simultaneously.
The field of view of the JCMT large PI project is shown in green solid lines in Figure \ref{fig:m31_obs}.
All the observations were reduced using the JCMT's standard \textsc{Starlink} software\footnote{https://starlink.eao.hawaii.edu/starlink}.
We reduced the individual observations using the {\tt makemap} algorithm in SMURF \citep{Chapin2013MNRAS} and then mosaicked them together.
We used a pixel size of 2$''$ and 4$''$ at 450 and 850 $\mu$m with the final map (angular resolutions of 8$''$ at 450 $\mu$m and 14$''$ at 850 $\mu$m), which reached a sensitivity of $\sim$31.6 and 1.2 mJy\,beam$^{-1}$, respectively.
Compared to the HASHTAG survey \citep{Li2020MNRAS,SmithApJS,Deng2025MNRAS}, which mapped the entire M31 galaxy (see Figure \ref{fig:m31_obs}, the blue dot–dashed lines trace the field of view of the HASHTAG survey), our survey focuses specifically on the inter-arm regions within $\sim$15 kpc and achieves $\sim$3 times better sensitivity at 850 $\mu$m. 
We incorporated HASHTAG SCUBA-2 data into our final maps for the overlapping field of view, which were subsequently used for image combination and SED analysis.

\subsection{Supplementary Data}\label{section:obs_sup}

M31 has been mapped using the PACS instrument \cite{Poglitsch2010A&A}, the SPIRE instrument \cite{Griffin2010A&A} on {\it Herschel}, and the High Frequency Instrument \citep{Lamarre2010A&A} on {\it Planck}.
The following analysis uses 100/160/250/350/500 $\mu$m {\it Herschel} data from \cite{Fritz2012A&A,Groves2012MNRAS}.  
The {\it Herschel} maps have an angular resolution of 12.5$''$, 13.3 $''$, 18.2$''$, 24.5$''$, and 36.0$''$ at 100, 160, 250, 350, and 500 $\mu$m, respectively.
We retrieved the {\it Planck} High Frequency Instrument 353 GHz image\footnote{Based on observations obtained with {\it Planck} (http://www.esa.int/Planck), an ESA science mission with instruments and contributions directly funded by ESA Member States, NASA, and Canada.}, which is presented in \cite{Planck2015A&A}.

To investigate the molecular hydrogen gas content in M31, we used $^{12}$CO J=1-0 observations taken by the IRAM 30 m telescope \citep{Nieten2006A&A}.
This was taken during 1995-2001, covering an area of about 2.0 $\deg$ $\times$ 0.5 $\deg$ (extending to $\sim$ 18 kpc) with a velocity resolution of 2.6 km s$^{-1}$ and an angular resolution of 23$''$ (see details in Section \ref{subsubsec:co}).
The 1$\sigma$ noise level in the velocity-integrated map is around 0.35 K km s$^{-1}$ \citep{Nieten2006A&A}.

\subsection{Combining JCMT-SCUBA2 Images with Herschel and Planck Observations}\label{section:obs_com}

Ground-based bolometric (sub)millimeter continuum mapping observations on spatially extended target sources are often subject to significant missing fluxes \citep{Chapin2013MNRAS.430.2545C}. 
This hampers accurate quantitative analyses. 
Missing flux can be recovered by fusing high-resolution images with observations that preserve extended structures (\citealt{Liu2015ApJ...804...37L}).
In this work, we combined our SCUBA2 images with {\it Herschel} \citep{Fritz2012A&A,Smith2012ApJ,Draine2014ApJ} and {\it Planck} \citep{Planck2015A&A} archive images to recover the missing flux at large scale by using the J-comb algorithm \citep{Jiao2022SCPMA} following the similar procedure introduced in \cite{Lin2016ApJ,Lin2017ApJ,Jiao2022SCPMA}. 

For the 450 $\mu$m Band, we made interpolations to the 450 $\mu$m images, based on the modified blackbody SED fitting to the {\it Herschel} PACS 100/160 $\mu$m and SPIRE 250/350/500 $\mu$m images. 
We pre-convolved all images for the SED fitting to the same angular resolution of the SPIRE 500 $\mu$m image ($\sim$37$''$) and re-gridded these images with the same pixel size of the PACS 70 $\mu$m image (3.2$''$).
We linearly combined the SCUBA2 450 $\mu$m and the interpolated 450 $\mu$m image in the Fourier domain.
In this way, the combined 450 $\mu$m map has the same angular resolution as the original SCUBA2 image (8.0$''$) but recovers the extended emission structures.

For the 850 $\mu$m Band, the spatial frequency coverage of the JCMT-SCUBA2 observations does not overlap well with the {\it Planck} images.
To address this, we applied an iterative deconvolution procedure using the Lucy–Richardson algorithm \citep{lucy1974}.
The Lucy–Richardson algorithm guarantees that every pixel of the deconvolved image has a positive value, and the total flux of the deconvolved image is conserved.  
Following a similar process as the 450 $\mu$m image, we made interpolations to the 850 $\mu$m image, based on the modified blackbody SED fitting to the PACS 100/160 $\mu$m and the SPIRE 250/350/500 $\mu$m images.
In the first step, we used this extrapolated {\it Herschel} 850 $\mu$m image as the model image in our deconvolution procedure applied to {\it Planck} 353 GHz image.
We convolved the deconvolved {\it Planck} image to a resolution of 60.0$''$ and combined it with the SCUBA2 850 $\mu$m image in the Fourier domain by using the J-comb algorithm.
In the second step, the combined SCUBA2 and {\it Planck} image was adopted as an improved model for a further deconvolution of the original {\it Planck} image, refining the correction of spatial frequencies.
We obtained a further deconvolved {\it Planck} image with a resolution of 60.0$''$ after the iteration.
After these, we combined the final deconvolved {\it Planck} image with the SCUBA2 image in the Fourier domain to obtain the combined map, which has an angular resolution of 14.0$''$.
In principle, this deconvolution procedure could be iterated further, but we found that two rounds of deconvolution were sufficient to maintain overlapping spatial frequencies for JCMT-SCUBA2 and {\it Planck} observations at 850 $\mu$m Band.
The central panel of Figure \ref{fig:scuba2_850} shows the stacked and combined 850 $\mu$m image, while the orange dashed rectangles highlight eight zoomed-in, selected regions (A-H), to present detailed substructures.
The stacked and combined 450 $\mu$m image is shown in Figure \ref{fig:scuba2_450}.

\begin{figure*}[]
\centering
\includegraphics[width=0.327\linewidth]{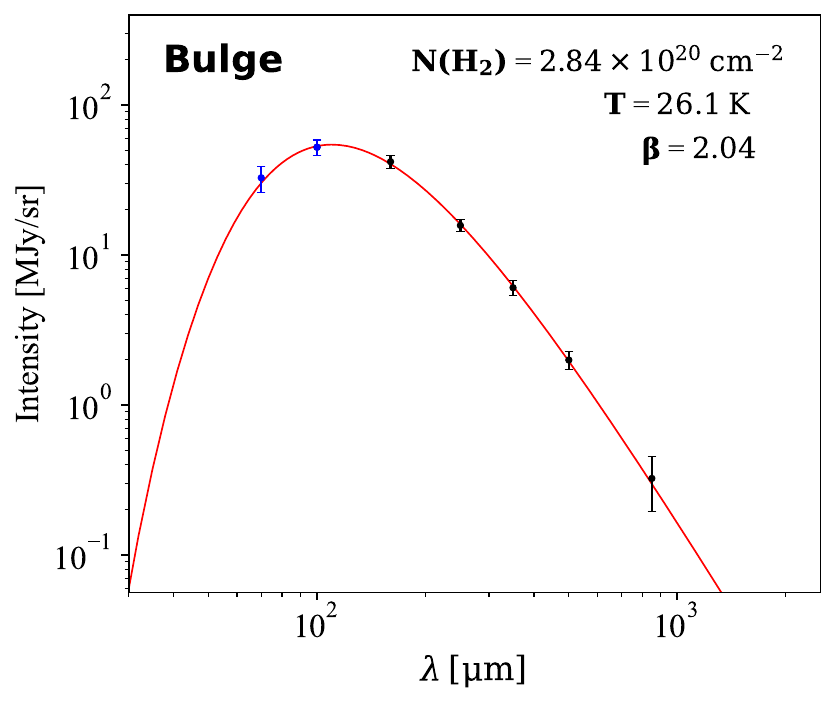} 
\includegraphics[width=0.327\linewidth]{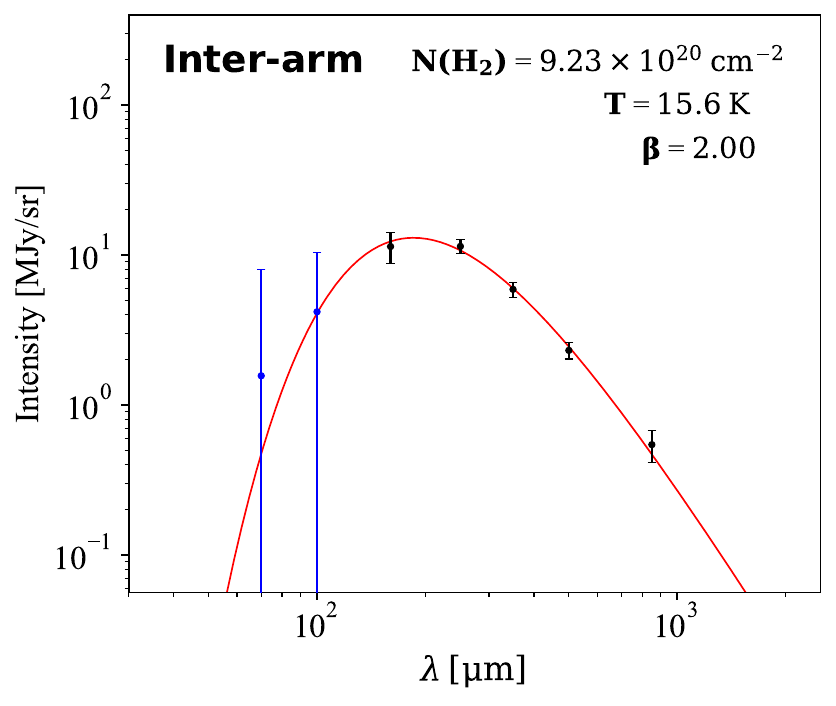} 
\includegraphics[width=0.327\linewidth]{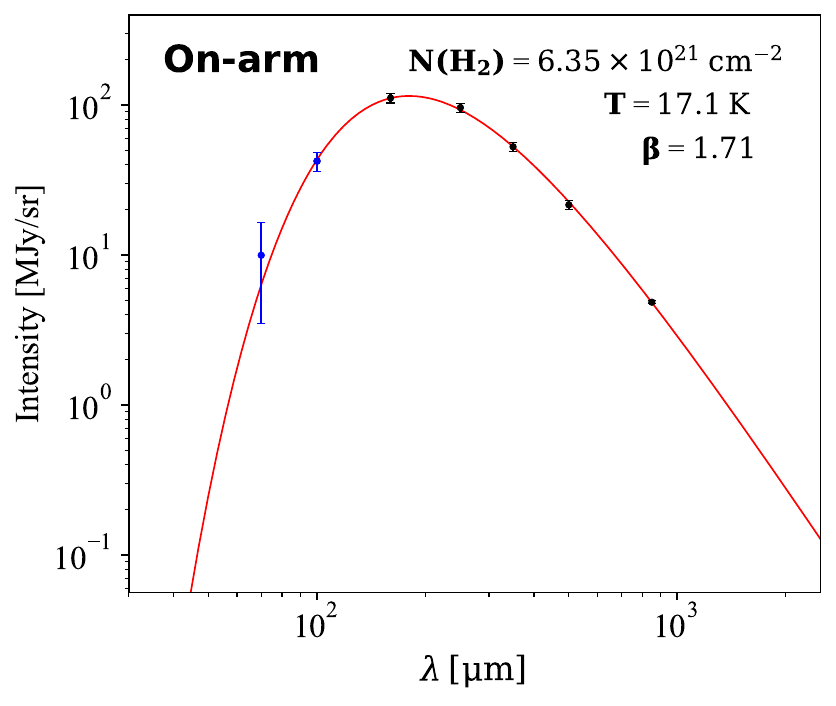}\\
\caption{
SEDs for pixels in different regions, constructed using  {\it Herschel} 70/100/160/250/350/500 $\mu$m and JCMT-SCUBA2 850 $\mu$m measurements.
The central pixel coordinates (RA/Dec) for each region are: 00:42:40.02; +41:17:29.36 (bulge region), 00:42:38.87; +41:27:35.34 (inter-arm region), and 00:41:05.48; +40:38:19.38 (on-arm region).
The 70 and 100 $\mu$m measurements (blue points) are used as upper limits in the SED fitting. 
}
\label{fig:sed_example}
\end{figure*}

\subsection{SED fitting}\label{section:obs_sed}

We have performed single-component, modified black body SED fitting to each pixel of the input images.
Before performing any SED fitting, we convolved all images to a common angular resolution of the largest telescope beam, and all images were re-gridded to have the same pixel size of 6$''$ per pixel.
In the least-squares fits, we weighted the data points by the noise level.
As modified black-body assumption, the flux density $S_{\nu}$ at a certain observing frequency $\nu$ is given by
\begin{equation}\label{eq:sed_1}
S_{\nu} = \Omega_{m}B_{\nu}(T_{\mbox{\scriptsize dust}})(1-e^{-\tau_{\nu}}),
\end{equation}
where the dust column density $N_{\mbox{\scriptsize dust}}$ can be approximated by
\begin{equation}\label{eq:sed_2}
N_{\mbox{\scriptsize dust}}=\tau_{\nu}/\kappa_{\nu},
\end{equation}
$B_{\nu}(T_{\mbox{\scriptsize dust}})$ is the Planck function at temperature $T_{\mbox{\scriptsize dust}}$, $\Omega_{m}$ is the solid angle, the dust opacity $\kappa_{\nu}=\kappa_{\mbox{\tiny{857 GHz}}}(\nu/857\,GHz)^{\beta}$, $\beta$ is the dust emissivity index.
We adopted the dust opacity per unit mass at 857 GHz of 0.192 m$^{2}$ kg$^{-1}$ \citep{Draine2003ARA&A}.
Adopting the dust-to-gas ratio of 0.01 and a dust temperature of 17.5\,K, 3 $\sigma$ limit at 850\,$\mu$m band corresponds to mass of 3.5$\times$10$^4\,M_{\odot}$ for a point-source.
Examples of SED fits for individual pixels in the center (bulge) region, inter-arm region, and arm region are shown in Figure \ref{fig:sed_example}.

\begin{figure*}
\begin{tabular}{c}
\hspace{-0.45cm}\includegraphics[width=18.13cm]{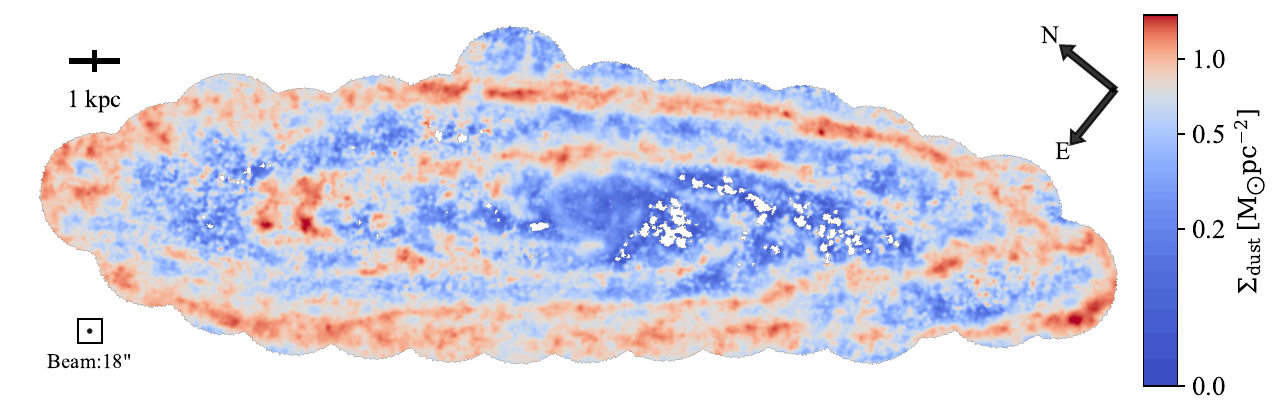} \\
\end{tabular}
\begin{tabular}{c}
\includegraphics[width=18.cm]{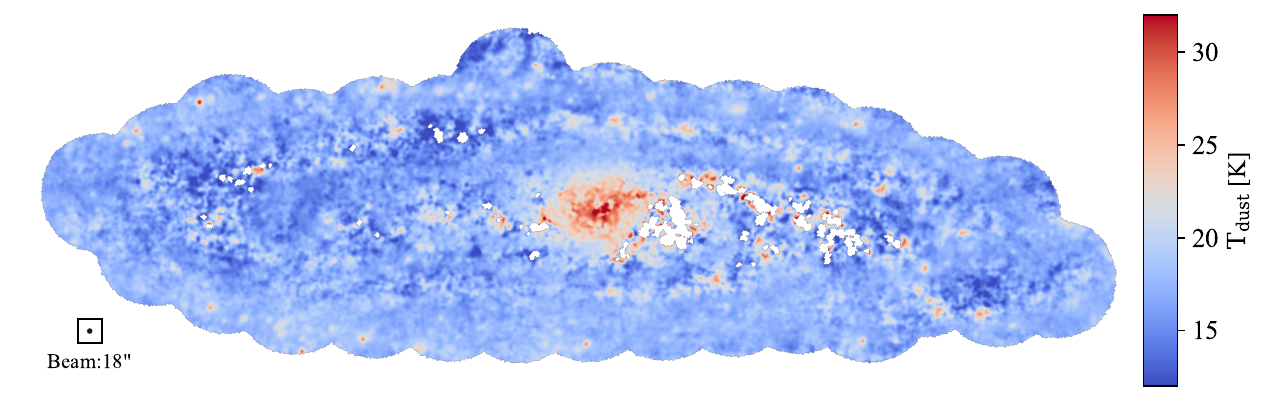} \\
\end{tabular}
\begin{tabular}{c}
\hspace{-0.05cm}\includegraphics[width=17.92cm]{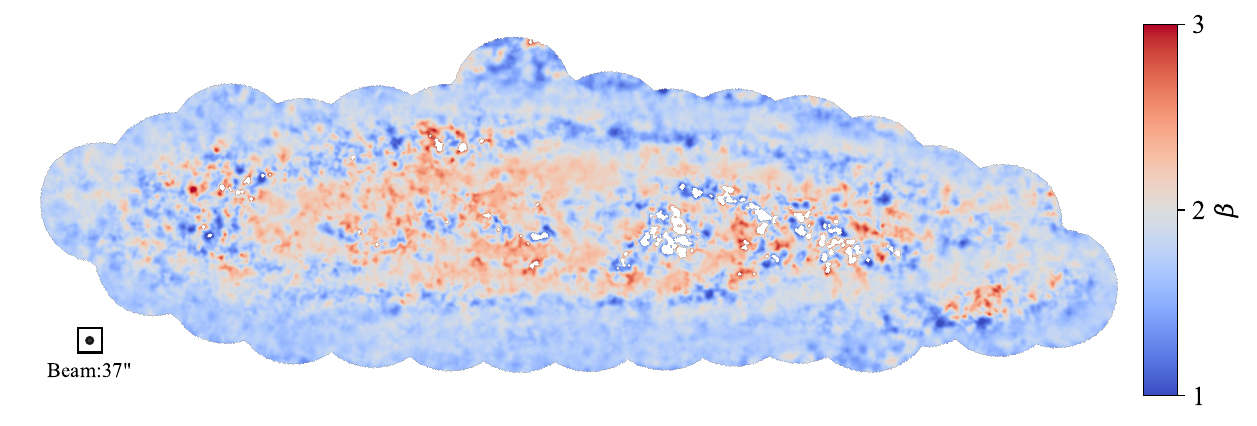} \\
\end{tabular}
\caption{
The distribution of dust surface density (18$''$ angular resolution), temperature (18$''$ angular resolution), and $\beta$ (37$''$ angular resolution) obtained from the iterative SED fits.
The detailed procedures are described in Section \ref{section:res_t_n}.
}
\label{fig:sed_maps}
\end{figure*}

Our procedure for iteratively deriving high-angular-resolution dust temperature and dust/gas column density images follows the methodology outlined in \cite{Lin2016ApJ,Lin2017ApJ}.
During the iterative SED fits, we first used 100, 160 $\mu$m from PACS, 250, 350, 500 $\mu$m from SPIRE, combined JCMT-SCUBA2 450, 850 $\mu$m data to simultaneously fit dust column density, dust temperature, and $\beta$.
In this step, the input images were convolved to a common angular resolution of 37$''$, which is slightly larger than the measured beam size of the SPIRE 500 $\mu$m image.
Additionally, pixels with fluxes in less than three bands that have a signal-to-noise ratio greater than 3$\sigma$ are masked.
The resulting 37$''$ resolution images of $\beta$ (see the bottom panel of Figure \ref{fig:sed_maps}), dust column density, and dust temperature are then used as initial values for the second iteration of the fitting process to aid in improving convergence.
During the second iteration, we used the values of $\beta$ from the last iteration and only fit to the 100, 160 $\mu$m from PACS, 250 $\mu$m from SPIRE, and combined JCMT-SCUBA2 450, 850 $\mu$m data.
This iteration yields the dust column density and dust temperature images with an angular resolution of $\sim$18$''$.

To estimate the uncertainties in our fitting results, we employed a Monte Carlo technique as described in \cite{Smith2012ApJ}.
For each pixel, we generated a set of 5000 artificial SEDs by adding random noise to the original flux densities. 
The noise was drawn from a normal distribution with a mean of zero and a standard deviation equal to the measurement uncertainty of the flux. 
We then perform SED fit for each pixel to estimate the uncertainties using the images at the same bands.
For $\beta$, we used artificially generated flux densities at 100/160/250/350/450/500/850 $\mu$m and found that the uncertainty in $\beta$ for each pixel is $\pm$0.27.
For dust column density and dust temperature, we used artificial flux densities at 100/160/450/850 $\mu$m. 
The uncertainties were found to be 20\% for the dust column density and $\pm$1.2 K for the dust temperature in each pixel.

\section{Spatial Distribution of Dust  Properties}\label{section:dust_map}

\begin{figure}
\vspace{0.1cm}
\includegraphics[width=9.cm]{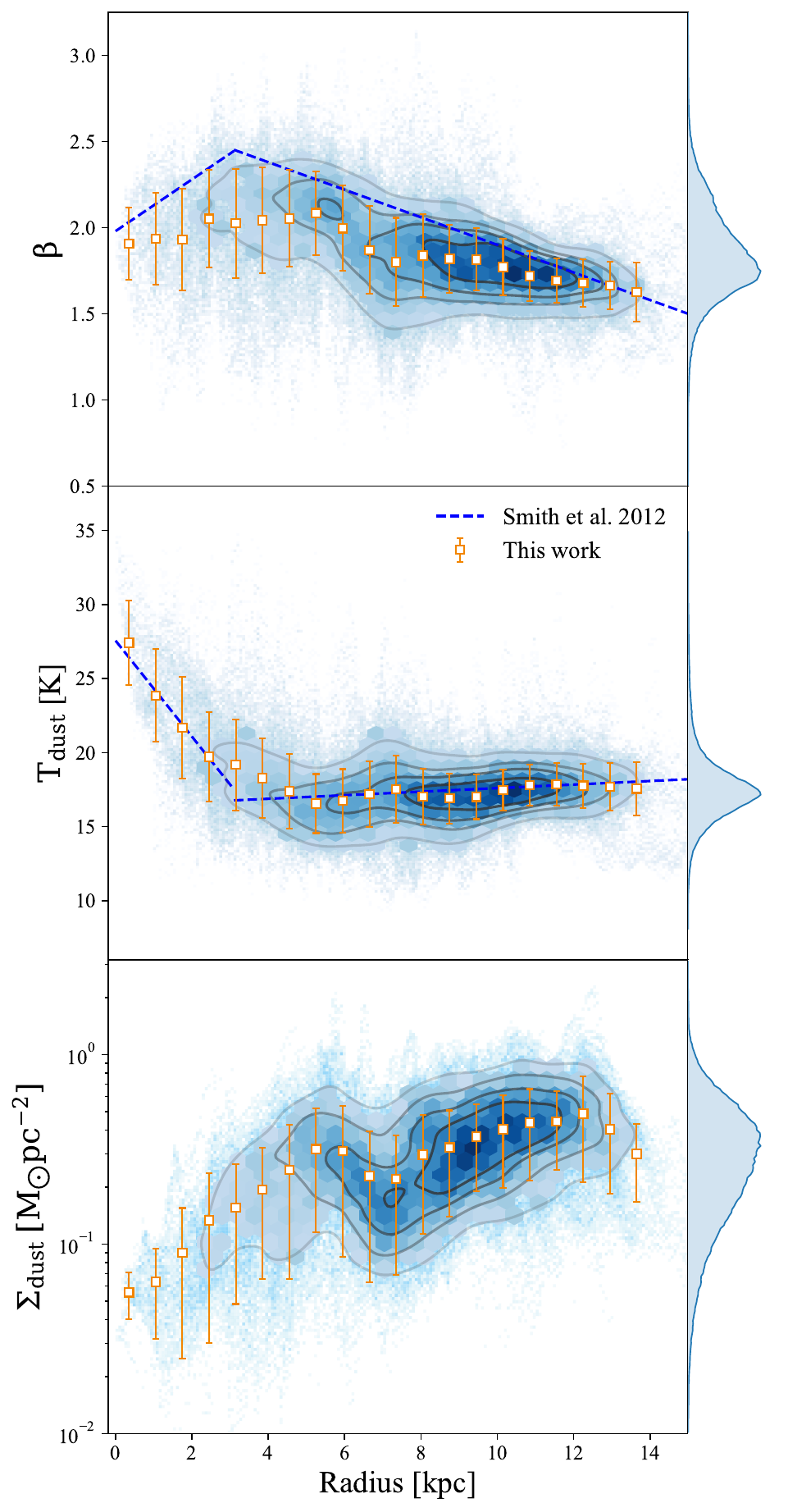}
\caption{Results from the SED fits for each pixel plotted vs. radius.
The blue points show the fitted value for each pixel, while the yellow squares represent averages in 700 pc bins.
The dashed blue lines represent the fitted relationships from the HELGA II \citep{Smith2012ApJ}.
The histograms on the right side show the dust surface density, dust temperature, and $\beta$ distribution of the whole map.
}
\label{fig:ori_sed_high_nh2}
\end{figure}

\subsection{Column Density, Temperature, and Dust Opacity Index Distribution Maps}\label{section:res_t_n}

Figure~\ref{fig:sed_maps} presents the results of our iterative SED fitting procedure: the dust surface density ($\Sigma_{\mbox{\scriptsize dust}}$) map at 18$''$ resolution, the dust temperature ($T_{\mbox{\scriptsize dust}}$) map at 18$''$ resolution, and the dust emissivity index ($\beta$) map at 37$''$ resolution.


The dust column density map (top panel of Figure~\ref{fig:sed_maps}) is broadly consistent with previous {\it Herschel}-based studies of M31 \citep[e.g.,][]{Smith2012ApJ,Draine2014ApJ}.
However, at 18$''$ resolution, our map resolves clustered and filamentary substructures along the spiral arms in much greater detail.
This higher level of detail is particularly valuable for analyses of the hierarchical morphology of spiral galaxies and potentially can be directly compared with large-scale giant molecular filaments or the so-called “bone” structures identified in the Milky Way \citep[e.g.,][]{Goodman2014ApJ,Ragan2014,Wang2015MNRAS,Abreu-Vicente2016,Zucker2018ApJ}.
In addition, our column density map newly reveals a number of GMCs in the inter-arm regions.

The dust temperature map (middle panel of Figure~\ref{fig:sed_maps}) is likewise consistent with previous {\it Herschel} results on large scales \citep[e.g.,][]{Smith2012ApJ}.
As expected, the highest dust temperatures are found in the bulge region, while there are dust temperature enhancements in the star-forming regions along the spiral arms.
For the outer disk and inter-arm regions, the dust temperatures are systematically lower.

In contrast, the spatial distribution of $\beta$ (bottom panel of Figure~\ref{fig:sed_maps}) shows notable differences from \citet{Smith2012ApJ}.
By including longer-wavelength data in our SED fitting, we obtain stronger constraints on $\beta$ than when using only the {\it Herschel} bands, leading to a relatively smooth and uniform distribution across the disk.
Although $\beta$ values within $R \lesssim 7$ kpc are generally higher than those at larger radii, our map does not exhibit the distinct high-$\beta$ ring ($\beta \sim 2.5$ at $R \sim 3$ kpc) reported by \citet{Smith2012ApJ}.

\subsection{Dust Property Trends with Galactocentric Radius}

Figure \ref{fig:ori_sed_high_nh2} shows the dust column density, temperature, and $\beta$ as functions of galactocentric radius, from the {\it Herschel} Exploitation of Local Galaxy Andromeda (HELGA) II \citep{Smith2012ApJ}.
In summary, these parameters are more uniform in the outer regions (e.g., R $\gtrsim$ 3 kpc), whereas the inner galaxy shows a lower gas column density, higher dust temperature, and higher $\beta$. 

The relationship between dust/gas properties and galactocentric radius has been explored in previous studies \citep[e.g.,][]{Smith2012ApJ,Draine2014ApJ,Whitworth2019MNRAS}.
Earlier {\it Herschel} observations reported significant radial variations in dust temperature and $\beta$ within Andromeda's disk \citep[e.g., HELGA II and V,][]{Smith2012ApJ,Mattsson2014}.
Comparing our measurements to those from HELGA II, we find consistent trends in dust temperature with galactocentric radius, but differences in the behavior of $\beta$.

\cite{Smith2012ApJ} identified a sharp change in $\beta$ at R$\approx$3 kpc, with $\beta$ rising from 2 to 2.5 within 3\,kpc and then decreasing from 2.5 to 1.5 from $\sim$3\,kpc to the $\sim$ 15 kpc radii.
Using the SPIRE500/SPIRE250 band ratio to measure $\beta$, \cite{Draine2014ApJ} observed distinct behavior for R $\gtrsim$ 7 kpc, finding the dust at R  $\gtrsim$ 7 kpc has $\beta$ $\approx$ 2.08, consistent with the dust model in the \cite{Draine2007ApJ}. 

The top panel of Figure \ref{fig:ori_sed_high_nh2} illustrates the variations in $\beta$ as a function of galactocentric radius, derived from modified black body SED fitting using {\it Herschel} 100/160/250/350/500 $\mu$m, and the combined JCMT-SCUBA2 450/850 $\mu$m images, with galacto-centric radius.
The inclusion of longer wavelength data in the SED fitting provides stronger constraints on $\beta$ compared to using only {\it Herschel} bands and results in smaller variations in $\beta$ (1.5 $\lesssim$ $\beta$ $\lesssim$ 2.3), especially for R$\lesssim$7\,kpc.
Our results show an average $\beta$ value of 1.8, which is smaller than values reported in other studies, such as, \cite{Draine2014ApJ}, where $\beta$ was defined in a post-processing step, between 250 and 500 $\mu$m, and \cite{Whitworth2019MNRAS}, where $\beta$ was treated as an intrinsic parameter in the \textsc{PPMAP} \citep{Marsh2015MNRAS} analysis.
In future works, investigate dust property variations across different regions in detail, extending beyond radial trends.

\subsection{The Structure of Spiral Arms}\label{sec:arm}

M31's proximity has facilitated high-resolution observations across various wavelengths, yielding extensive data from infrared to submillimeter bands by e.g., {\it Spitzer} \citep{Gordon2006ApJ,Barmby2006ApJ}, {\it Herschel} \citep{Fritz2012A&A,Smith2012ApJ}, and {\it Planck} space telescopes \citep{Planck2015A&A}, as well as {\it GALEX} in the near and far UV band \citep{Thilker2005ApJ,ford2013ApJ}.
The spiral arms of M31 have been analyzed by using optical light \citep{Hodge1981}, HI radio line emission \citep{Chemin2009ApJ,Corbelli2010A&A}, CO line emission \citep{Nieten2006A&A}, infrared and submillimeter continuum emission \citep{Gordon2006ApJ,Kirk15,Tenjes2017A&A}.
Despite its classification as a de Vaucouleurs type SA(s)b \citep{Hubble1929ApJ}, M31's large-scale gas structures can not be conclusively described by several spirals alone.
Instead, the spiral structure has been approximated by a composite of two logarithmic spiral arms \citep{Braun1991ApJ,Nieten2006A&A,Gordon2006ApJ} and a circular ring (radius $\sim$10 kpc) of star formation offset from the nucleus \citep{Gordon2006ApJ}.
An additional circular ring (radius $\sim$15 kpc) has been added and a more complicated spiral structure has been considered with {\it Herschel} images \citep{Kirk15}.

\begin{figure*}[ht!]
\vspace{-0.3cm}
\hspace{0.5cm}
\includegraphics[width=17cm,]{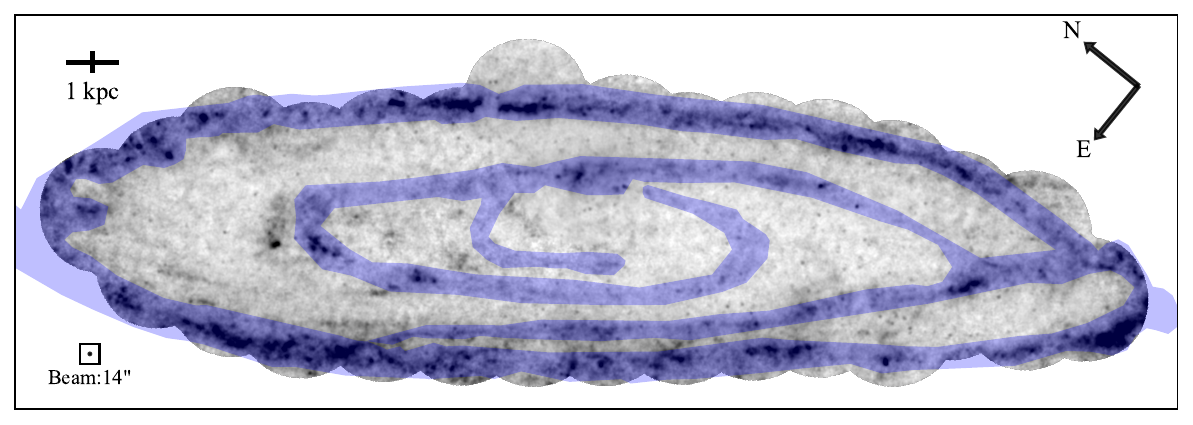}
\caption{
M31's arm mask is overlaid on the background of the combined 850 $\mu$m JCMT continuum image. 
The arm mask is highlighted by the blue shadow using the methodology discussed by \cite{Querejeta2021A&A...656A.133Q}.
Areas within the shadow are designated as the arm environment, while areas outside are designated as the inter-arm environment.
}
\label{fig:spiral_arm}
\end{figure*}

In this work, we divide the JCMT field of view into two environments: the arm environment and the inter-arm environment.
Due to M31's large inclination angle (i $\approx$ 77$^{\circ}$), it is unsuitable for using the Fourier decomposition method \citep{Yu2018ApJ...862...13Y, Chen2021ApJ} to quantify the main characteristics of galaxy spiral arms.
Here, we construct the arm mask for M31 following the general method introduced by \cite{Querejeta2021A&A...656A.133Q}, though applied here to dust tracers rather than stellar-mass tracers.
This arm mask delimits morphological features visually identified on infrared and submillimeter images, including spiral arms and rings. 

Specifically, the construction of the arm mask involved three key steps: 
first, defining the arm mask with a classic two-arm logarithmic spiral and two offset rings, following \cite{Kirk15}; second, assigning arm curves an empirically determined width based on dust tracers, specifically the {\it Spitzer} 24 $\mu$m band \citep{Gordon2006ApJ} and the {\it Herschel} 250 $\mu$m continuum \citep{Smith2012ApJ,Draine2014ApJ}; and finally, visually inspecting and adjusting the resulting masks to ensure continuity. 
Figure \ref{fig:spiral_arm} displays the combined 850 $\mu$m JCMT continuum image with the overlaid arm mask, which is highlighted in blue shadow. 
Areas within the blue shadow are designated as the arm environment, while areas outside are considered as the inter-arm environment.
Using this arm mask and assuming a distance of 766$\pm$18 kpc \citep{Lee2023ApJ}, the inter-arm dust mass in the field of view is estimated to be 1.0$\times$10$^7$ $M_{\odot}$, accounting for 37\% of the total dust mass in our mapped region. 
Adopting a total dust mass of 5.4$\times$10$^7\,M_{\odot}$ for M31 from \cite{Draine2014ApJ}, we find that approximately 19\% of M31's dust is located in the inter-arm region within the 10 kpc ring.

\section{Gaint Molecular Clouds}\label{sec:gmcs}

In this section, we discuss the identification of the molecular clouds in M31 and the corresponding measurements for each cloud. 
These quantities, listed in Table \ref{table:quantities}, collectively form the cloud catalog produced by this study.

\subsection{Cloud Identification}

We identify and characterize the GMCs in M31 based on the 850 $\mu$m combined continuum image (Figure \ref{fig:scuba2_850}). 
Given the highly hierarchical emission and broad dynamical range of spatial scales, we employ the {\it Dendrogram} algorithm \citep{Rosolowsky08}, implemented with {\it astrodendro}, to carry out an automated and systematic search. 
The {\it Dendrogram} algorithm is widely used for identifying and characterizing hierarchical structures in the continuum or line data \citep[e.g.,][]{Kirk15,Ginsburg16,Cheng18}.

At the distance of M31, the angular resolution of the 850 $\mu$m image (14$''$) corresponds to a spatial resolution of $\sim$50 pc, which is comparable to the typical size of GMCs in the Milky Way \citep{Solomon1979ApJ}. 
To highlight GMC-scale structures, we apply the Constrained Diffusion Decomposition (CDD) technique \citep{Li2022ApJS} to the 14$''$ 850 $\mu$m {\it Planck}+JCMT-SCUBA2 image of M31 (the central panel of Figure \ref{fig:scuba2_850}) and separate it into components across multiple angular scales. 
We retain structures smaller than $\sim$300 pc in the decomposed image and analyze it using the {\it Dendrogram} algorithm.
We assume the identified leaves (the base element in the hierarchy of {\it Dendrogram} that has no further substructure) as individual GMCs.
We set the minimum flux density threshold to 4$\sigma$ of the decomposed image, the minimum significance for structures to 1$\sigma$, and the minimum area to the size of the beam (14\arcsec). 
Appendix \ref{gmc_example} presents detailed views of the identified sources for two representative subregions.
These settings give 572 GMCs, of which 189 are inter-arm clouds based on the arm mask introduced in Section \ref{sec:arm}. 
We have tested different parameter combinations, and this fiducial case ensures an optimized balance between maximizing detection completeness and minimizing false positives.

\subsection{Giant Molecular Cloud Properties from {\it Dendrogram}}

In the {\it Dendrogram} analysis, each cloud is characterized by four parameters [$C_i$, $A_i$, $\alpha_i$, $\delta_i$], which describe its cloud identification ($C_i$), physical area of the cloud ($A_i$), and RA/Dec position ([$\alpha_i$, $\delta_i$]), respectively.
The effective radius of a cloud is defined as
\begin{equation} \label{eq:R_c}
R_{\mbox{\scriptsize c}} = \sqrt{A/\pi}. 
\end{equation}
We adopt the cloud area returned by {\it Dendrogram}, which is defined with an isophotal boundary at a certain flux level, i.e., the level where two GMCs merge together or the 4$\sigma$ flux threshold for isolated GMC. 
So the cloud area or radius could be underestimated in a crowded field. 
It is assumed that a cloud's area is not affected by the projection of M31.

Considering the inclination angle i $\approx$ 77$^{\circ}$ of the M31 disk, the galatocentric radius $R$ can be calculated using the following equation:
\begin{equation}
    \begin{aligned}
        R = \sqrt{(d\cos{\phi})^2 + (\frac{d\sin{\phi}}{\cos{i}})^2},
    \end{aligned}
    \label{eq14}
\end{equation}
where $d$ is the projected distance from the center of M31, located at $\alpha$ = 10$^{\circ}$.685 and $\delta$= 41$^{\circ}$.269 \citep{Crane1992ApJ...390L...9C}, $\phi$ is the angle between the line connecting the cloud to the galactic center and the major axis of the M31 disk.

\begin{deluxetable}{l c l}
\tabletypesize{\scriptsize}
\tablenum{1}
\tablecaption{Quantities of the Molecular Cloud Catalog}\label{table:quantities}
\setlength{\tabcolsep}{20pt}
\tablewidth{5pt}
\tablehead{
\colhead{Quantity} & \colhead{Units} & \colhead{Description}
}
\startdata
$C$ & ... & Cloud index \\
$A$ & pc$^{2}$ & Physical area\\
$\alpha$ & deg  & Central RA \\
$\delta$ & deg  & Central Dec \\
$R_{\mbox{\scriptsize c}}$ &  pc  & Effective size \\
$R$ &  kpc  & Annular radius \\
$T_{\mbox{\scriptsize dust}}$ & K & Dust temperature \\
$\beta$ & ... & Dust opacity index \\
$\Sigma_{\mbox{\scriptsize dust}}$ & $M_{\odot}$ pc$^{-2}$ & Dust surface density   \\
$\Sigma_{\mbox{\scriptsize gas}}$ & $M_{\odot}$ pc$^{-2}$ & Gas surface density  \\
$M$ & $M_{\odot}$ & Mass  \\
$n_{\mbox{\scriptsize H}}$ & cm$^{-3}$ & Gas mean volume density  \\
\enddata
\end{deluxetable}

\begin{deluxetable*}{ccccccccccccc}
\tablecaption{Physical Properties of Selected Regions\label{table:GMC_info}}
\tablenum{2}
\tablehead{
\colhead{Index} & \colhead{On Arm} & \colhead{RA (J2000)} & \colhead{Dec (J2000)} & \colhead{$R_c$} & \colhead{$R$} & \colhead{$T_\mathrm{dust}$} & \colhead{$\beta$} & \colhead{$\Sigma_\mathrm{dust}$} & \colhead{$\Sigma_\mathrm{gas}$} & \colhead{Mass} & \colhead{$n_\mathrm{H}$} & \colhead{$\Sigma_\mathrm{SFR}$} \\
\colhead{} & \colhead{} & \colhead{} & \colhead{} & \colhead{(pc)} & \colhead{(pc)} & \colhead{(K)} & \colhead{} & \colhead{(M$_\odot$\,pc$^{-2}$)} & \colhead{(M$_\odot$\,pc$^{-2}$)} & \colhead{($10^5$\,M$_\odot$)} & \colhead{(cm$^{-3}$)} &
\colhead{(M$_\odot$\,yr$^{-1}$\,kpc$^{-2}$)} 
}
\startdata
1  & True  & 00:41:00.32  & 40:36:50.0  & 77.279  & 12.704 & 19.5 & 1.63 & 0.97 & 118.56 & 22.24 & 33.26 & 0.0026 \\
2  & True  & 00:40:55.74 & 40:37:21.0  & 118.559 & 12.049 & 17.7 & 1.76 & 1.14 & 135.46 & 59.82 & 24.77 & 0.0016 \\
3  & True  & 00:41:03.80  & 40:37:50.0  & 46.743  & 12.456 & 16.8 & 1.75 & 1.67 & 201.73 & 13.85 & 93.55 & 0.0013\\
4  & True  & 00:41:05.90  & 40:38:18.0  & 44.344  & 12.381 & 16.3 & 1.78 & 1.62 & 195.56 & 12.08 & 95.59 & 0.0010\\
5  & True  & 00:41:09.37  & 40:39:34.0  & 64.993  & 11.987 & 16.8 & 1.73 & 0.86 & 101.68 & 13.49 & 33.91 & 0.0006\\
6  & False & 00:40:42.32 & 40:39:16.0  & 31.931  & 10.328 & 17.8 & 1.60 & 0.34 & 36.34  & 1.16  & 24.67 & 0.0003\\
7  & False & 00:40:52.85 & 40:39:41.0  & 50.488  & 10.686 & 16.1 & 1.55 & 0.44 & 48.35  & 3.87  & 20.76 & 0.0003\\
8  & True  & 00:41:12.18 & 40:39:42.0  & 35.187  & 12.163 & 15.7 & 1.83 & 0.91 & 108.11 & 4.20  & 66.60 & 0.0007\\
9  & True  & 00:41:20.59 & 40:40:43.0  & 74.398  & 12.389 & 17.6 & 1.79 & 0.54 & 65.00  & 11.30 & 18.94 & 0.0007\\
10 & True  & 00:41:13.90 & 40:40:50.0  & 77.749  & 11.686 & 16.3 & 1.64 & 0.63 & 73.14  & 13.89 & 20.39 & 0.0005\\
\enddata
\tablecomments{This table is published in its entirety in the machine-readable format. The results shown here are for guidance regarding its form and content.}
\end{deluxetable*}

\subsection{Giant Molecular Cloud Properties from SED Fitting}

Assuming no significant variations across the resolved spatial scales of a cloud, the averaged dust temperature ($T_{\mbox{\scriptsize dust}}$) and averaged dust emissivity index ($\beta$) of each cloud are derived from the 18$''$ resolution dust temperature and 37$''$ resolution $\beta$ distribution maps (see \ref{section:res_t_n}).
The dust surface density ($\Sigma_{\mbox{\scriptsize dust}}$) of each cloud is calculated using single-component, modified blackbody fits based on $T_{\mbox{\scriptsize dust}}$, $\beta$, and the 850 $\mu$m intensity ($S_{\rm 850\mu m}$), as outlined in Equation \ref{eq:sed_1} and \ref{eq:sed_2}.
The detailed assumptions and calculations are introduced in \ref{section:obs_sed}.

To determine gas surface density for each cloud from its dust surface density, we need to use a dust-to-gas ratio ($g$).
Utilizing HI observations taken by WSRT \citep{Braun2009ApJ...695..937B}, CO observations taken by IRAM 30m \citep{Nieten2006A&A}, and dust observations taken by {\it Spitzer} \citep{Gordon2006ApJ} and {\it Herschel}, \cite{Draine2014ApJ} derived the dust-to-gas ratio map of M31 and revealed a decline in $g$ with increasing annular distance, $R$, from the galactic center:
\begin{equation}\label{eq:dust-to-gas}
g(R) \approx \begin{cases}
            0.0280 \, e^{-R/8.4 \, {\rm kpc}}  
            &(R < 8.4 \, {\rm kpc}) \\
            0.0165 \, e^{-R/19 \, {\rm kpc}}  
            &(8.4 \, {\rm kpc} < R < 18 \, {\rm kpc})\\
             0.0605 \, e^{-R/8 \, {\rm kpc}} 
            &(18 \, {\rm kpc} < R \lesssim 25 \, {\rm kpc}).
        \end{cases}
\end{equation}
Adopting $g(R)$ from \cite{Draine2014ApJ}, we computed the gas surface density for each cloud as follows:
\begin{equation}\label{eq:N_gas}
\Sigma_{\mbox{\scriptsize gas}} =  \Sigma_{\mbox{\scriptsize dust}} / g(R).
\end{equation}
The cloud mass ($M$) of each cloud can be calculated by summing up $\Sigma_{\mbox{\scriptsize gas}}$ from the cloud's physical area by
\begin{equation} \label{eq:M}
    M = A \, \Sigma_{\mbox{\scriptsize gas}}.
\end{equation}
Assuming spherical geometry, we calculated the gas mean volume density for each cloud:
\begin{equation} \label{eq:n}
    n_{\mbox{\scriptsize H}} = M / (\frac{4}{3} \pi R^{3}_{\rm c} \mu m_{\rm H}),
\end{equation}
where $\mu=2.8$ is the mean molecule weight in the interstellar medium \citep{Kauffmann2008}, $m_{\rm H}$ is the mass of a hydrogen atom.

\subsection{Distribution Functions of Physical Parameters}\label{sec:distribution}

\begin{figure*}[!ht]
\centering
\includegraphics[width=0.46\linewidth]{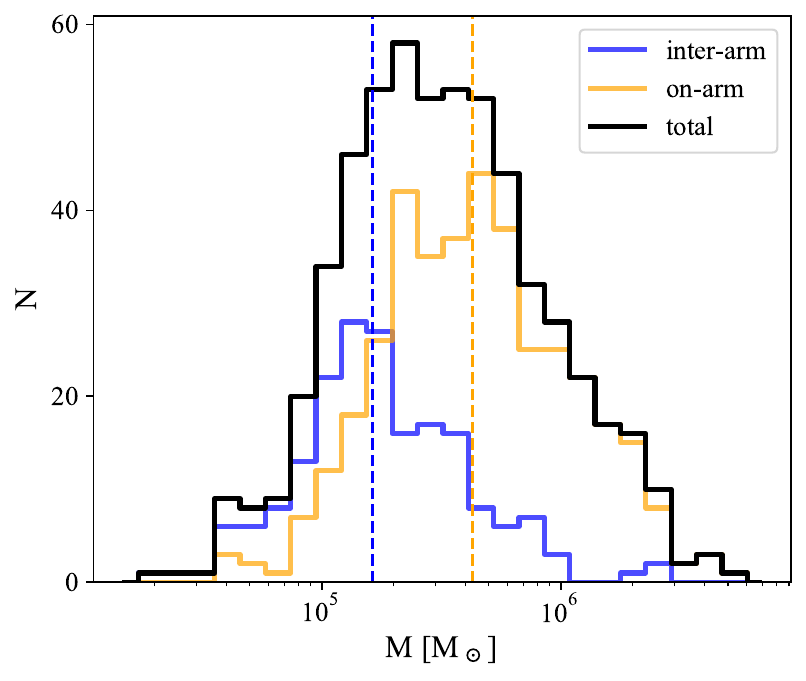} 
\includegraphics[width=0.46\linewidth]{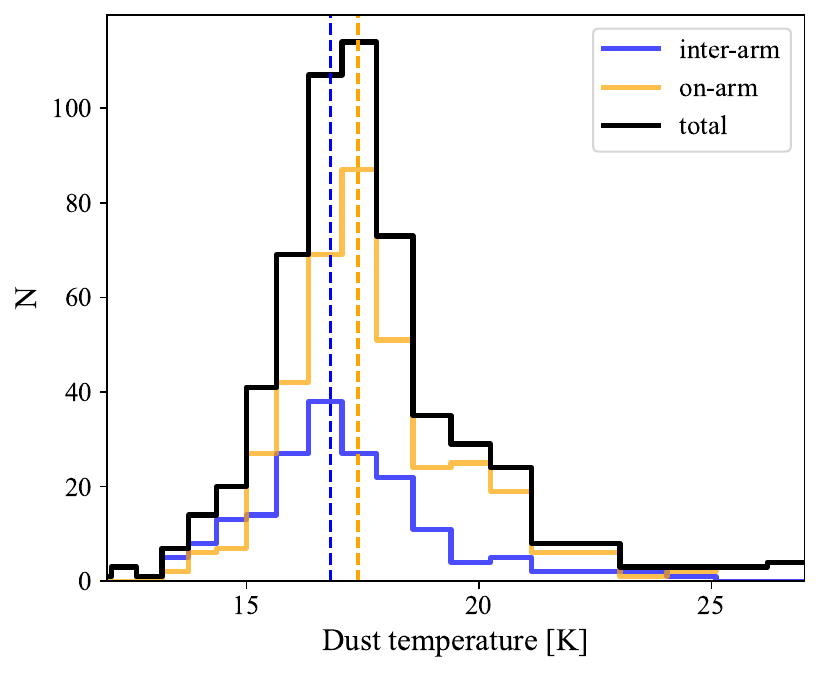} \\
\includegraphics[width=0.46\linewidth]{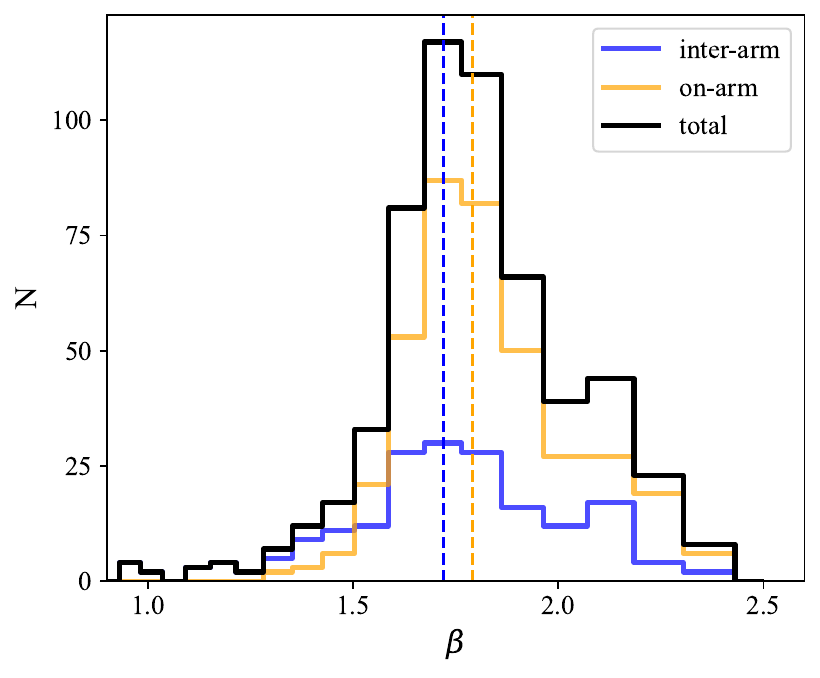}
\includegraphics[width=0.46\linewidth]{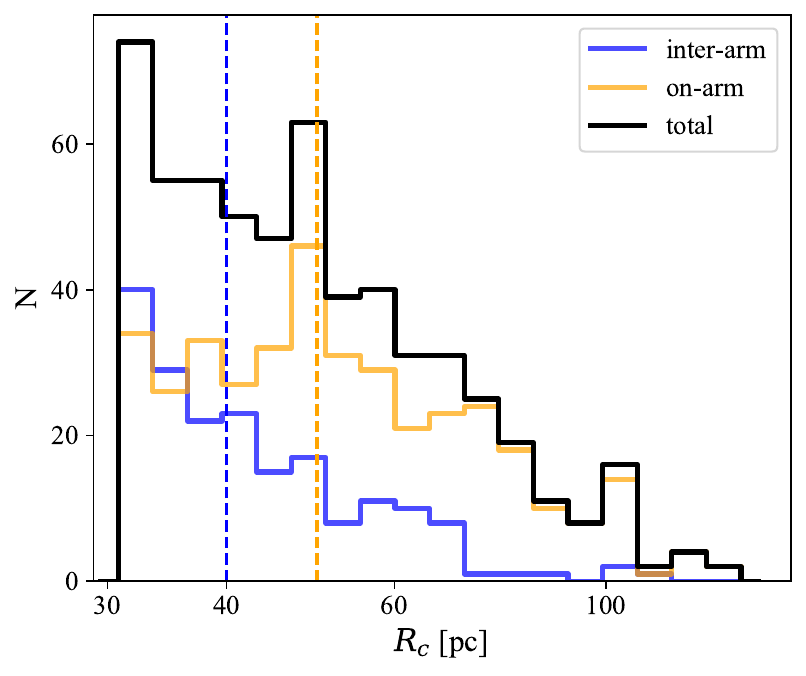} \\
\includegraphics[width=0.46\linewidth]{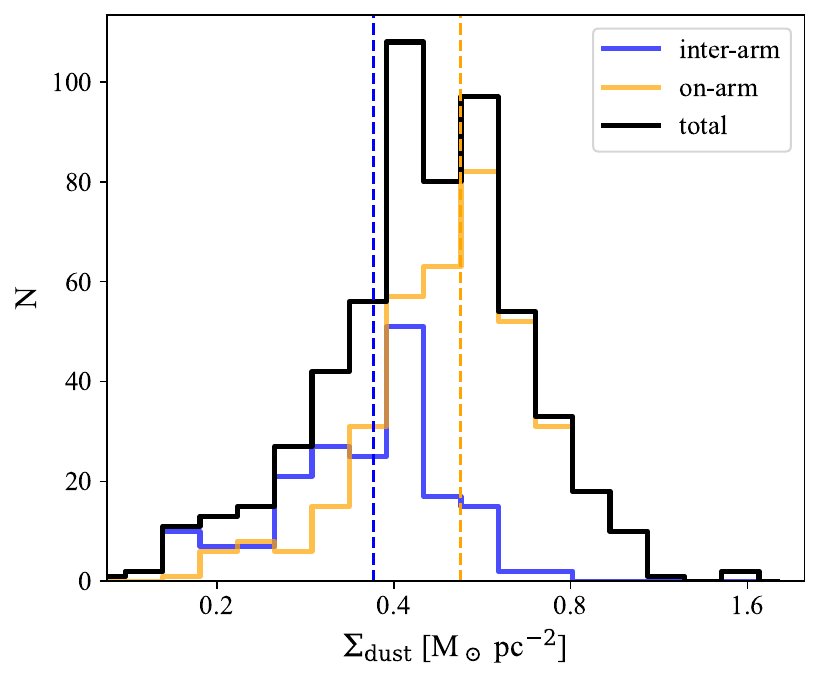} 
\includegraphics[width=0.46\linewidth]{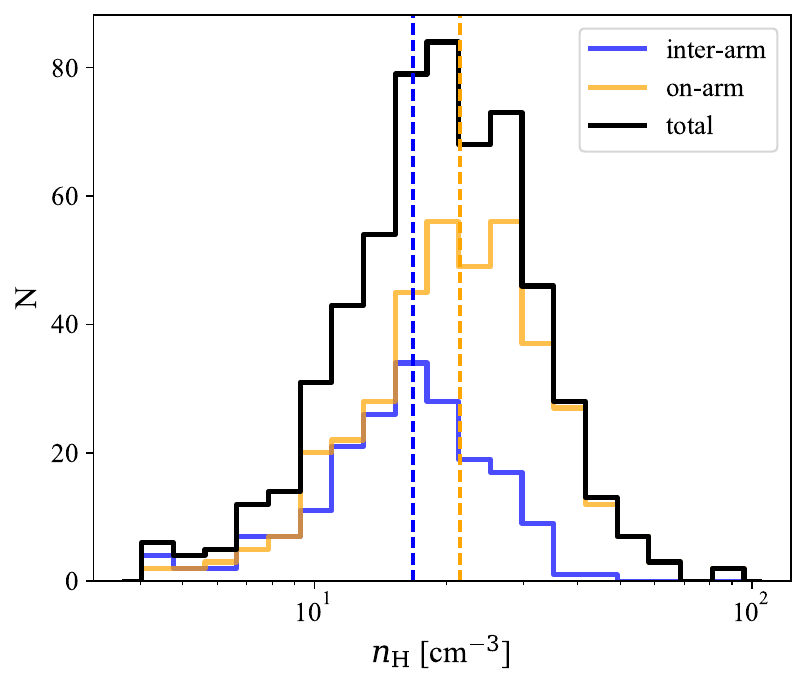}\\
\caption{
Probability distribution functions (histograms) of Mass, dust temperature, $\beta$, R$_{c}$, $\Sigma_{dust}$ and $n_{\mbox{\scriptsize H}}$ for all the clouds in the catalog (black histograms), for on-arm clouds (orange histograms), and for inter-arm clouds (blue histograms). 
The blue and orange dashed lines show the median values for off-arm clouds and on-arm clouds, respectively.
The typical values for these quantities are summarized in Table \ref{tab:sum}.
}
\label{fig:distribution}
\end{figure*}

The distributions of cloud Mass, dust temperature, $\beta$, R$_{c}$, gas surface density, and gas volume density for all the clouds are shown in Figure \ref{fig:distribution}.
All the typical values of these cloud parameters are tabulated in Table \ref{tab:sum}.

The majority of cloud has a mass range from 4.0$\times$10$^4$ $M_{\odot}$ to 3.0 $\times$ 10$^6$ $M_{\odot}$, while the median mass of the sample is 3.0 $\times$ 10$^5$ $M_{\odot}$.
This is comparable to the molecular masses of several well-known star-forming regions \citep[e.g.,][]{Lin2016ApJ...828...32L}, including W51 \citep{Carpenter1998AJ} and the Carina Nebula complex \cite{Preibisch2011A&A}.
Notably, there is a significant difference between the cloud mass distributions of the on-arm and inter-arm GMCs in M31 (Figure \ref{fig:distribution}).
Clouds in the arm regions are significantly more massive, with a median mass of 4.3 $\times$ 10$^5$ $M_{\odot}$, compared to $1.6 \times 10^5 M_{\odot}$ for the inter-arm GMCs.

\begin{deluxetable*}{cccccccccc}[!ht]
\tabletypesize{\scriptsize}
\tablenum{3}
\tablecaption{Typical Values of Cloud Parameters}
\tablehead{
\colhead{Parameter} & \colhead{Units} & \multicolumn{2}{c}{All Clouds} & \multicolumn{2}{c}{On-arm Clouds} & \multicolumn{2}{c}{Inter-arm Clouds} & \multicolumn{2}{c}{MW Clouds}\\
\cmidrule(r){3-4} \cmidrule(r){5-6} \cmidrule(r){7-8} \cmidrule(r){9-10} 
 & & \multicolumn{1}{c}{Mean} & \multicolumn{1}{c}{Median} & \multicolumn{1}{c}{Mean} & \multicolumn{1}{c}{Median} & \multicolumn{1}{c}{Mean} & \multicolumn{1}{c}{Median} & \multicolumn{1}{c}{Mean} & \multicolumn{1}{c}{Median}
}
\decimals
\startdata
$M$ &10$^4$ $M_{\odot}$ & 53.0 & 29.9 & 66.2 & 42.8 & 26.2 & 16.3 &  28.6 & 16.5 \\
$T_{\mbox{\scriptsize dust}}$ & K & 17.7 & 17.5 & 17.8 & 17.4 & 16.9 & 16.8  & ... & ... \\
$\beta$ & ... & 1.78 & 1.77 & 1.82 & 1.79 & 1.71 & 1.72  & ... & ... \\
$R_{\mbox{\scriptsize c}}$ & pc & 51.9 & 46.7 & 55.7& 49.7 & 44.4 & 40.0  & 31.5 & 25.1 \\
$\Sigma_{\mbox{\scriptsize dust}}$ & $M_{\odot}$ pc$^{-2}$ & 0.48 & 0.45 & 0.54 & 0.52 & 0.36 & 0.37   & 0.29  & 0.17 \\
$n_{\mbox{\scriptsize H}}$ & cm$^{-3}$ & 21.5 & 19.4 & 23.6 & 21.5 & 17.3 & 16.8  & 24.1 & 9.6 \\
\enddata
\tablecomments{Mean and median values of physical quantities for different subsets of the catalog. 
The mean and median values of the Milky Way clouds are cited from \cite{Miville2017ApJ}, where $\Sigma_{\mbox{\scriptsize dust}}$ is derived based on the surface density with an assumption of the fixed dust-to-gas ratio of 0.01.
\label{tab:sum}}
\end{deluxetable*}

In addition, the dust temperatures of the inter-arm GMCs are also slightly colder in general.
The top-right panel of Figure \ref{fig:distribution} shows a histogram of dust temperatures obtained from the SED fitting.
The majority of cloud dust temperatures range from 14.0 to 22.0 K, with a median value of 17.2 K, consistent with results based on {\it Herschel} observations \citep{Kirk15}.
While the dust temperature histograms for both on-arm and inter-arm clouds exhibit unimodal distributions, inter-arm clouds are systematically colder, with a median dust temperature of 16.8 K, compared to 17.4 K for on-arm clouds.
Additionally, there are much fewer inter-arm clouds with temperatures exceeding 20 K.
We performed a Mann-Whitney U test \citep{Mann47} to evaluate differences between the two populations statistically. 
It is a nonparametric test for the ranking between on-arm and inter-arm clouds. 
The null hypothesis that on-arm clouds have equal or lower dust temperatures than inter-arm clouds can be rejected with a confidence greater than 99.99\% ($p = 6.8 \times$ 10$^{-7}$). 
The lower dust temperatures of the inter-arm clouds may indicate reduced star-formation activity in these regions, aligning with star formation rate (SFR) measurements derived from the FUV and 24 $\mu$m emissions \citep{ford2013ApJ}.

The $\beta$ distributions for on-arm clouds, inter-arm clouds, and the full sample all exhibit single-peak distributions, primarily ranging from 1.30 to 2.40.
In addition, the median $\beta$ values are closely aligned: 1.79 for on-arm clouds, 1.72 for inter-arm clouds, and 1.77 for the full sample.
These values are consistent with the average measurement found toward the Galactic plane of the Milky Way, where $\beta$ $\sim$ 1.8 \citep[][]{Planck2011A&A}. 

The clouds in the catalog have physical sizes ($R_{c}$) ranging from 30 to about 130 pc (see Figure \ref{fig:distribution}, middle right panel), with a median value of 46.7 pc.
This median size is comparable to the typical size of GMCs in the Milky Way \citep[$\sim$50 pc,][]{Solomon1979ApJ} and the mean radii reported in GMC catalogs of the Milky Way \citep[e.g.,][]{Rice2016ApJ,Miville2017ApJ}.
The median $R_{c}$ values for the on-arm and inter-arm clouds are 49.7 pc and 40.0 pc, respectively, suggesting that on-arm clouds may be systematically larger.
However, the larger size of on-arm clouds could also be influenced by the overlap effect, as more clouds are located in the arm regions.
Notably, the measurement of GMC physical sizes is highly impacted by the resolution of the observations, especially for smaller clouds.

The median dust surface density ($\Sigma_{\mbox{\scriptsize dust}}$) of all the clouds in the catalog is 0.45 $M_\odot$ pc$^{-2}$.
The dust surface density of on-arm clouds is higher, with a mean value of 0.52 $M_\odot$, compared to 0.37 $M_\odot$ for inter-arm clouds.

The gas volume density ($n_{\mbox{\scriptsize H}}$) distributions are shown in the bottom right panel of Figure \ref{fig:distribution}, which are similar to $\Sigma_{\mbox{\scriptsize dust}}$ case.
The median volume density of the whole sample is 19.4 cm$^{-3}$.
The median values for the on-arm clouds and off-arm clouds are 21.5 and 16.8 cm$^{-3}$, respectively.
These typical values are close to the measurement for the GMCs in the Milky Way \citep{Miville2017ApJ}.
The difference in $n_{\mbox{\scriptsize H}}$ between on-arm and inter-arm clouds may suggest that inter-arm clouds lack dense gas. 
As a result, inter-arm clouds may not contain sufficient dense gases to support significantly massive star formation, which is consistent with the lower dust temperatures observed in this sample.

\subsection{Mass Spectrum}

\begin{figure}
\includegraphics[width=1.03\linewidth]{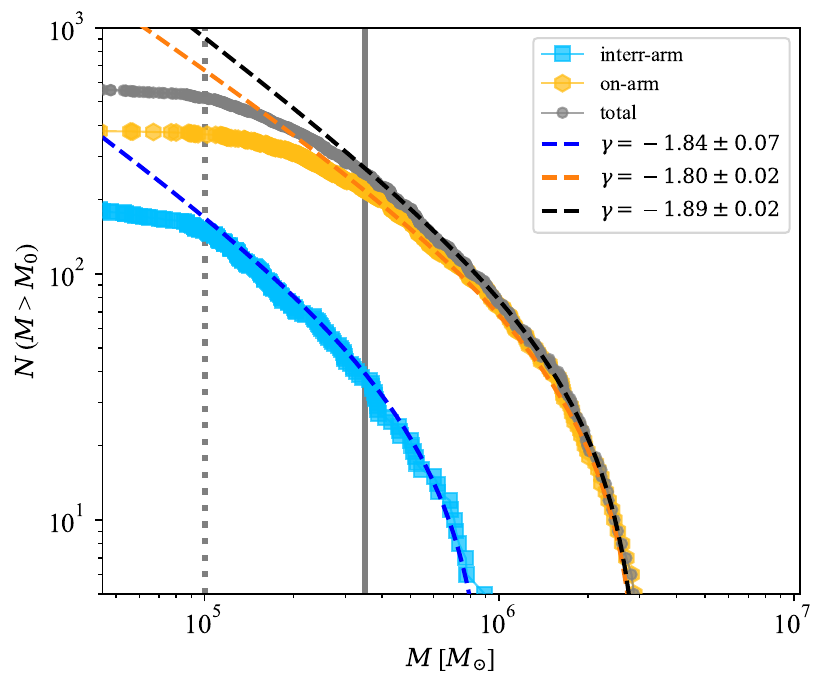}
\caption{
Cumulative mass spectra for the full sample (gray points), for on-arm clouds (orange hexagons), and for inter-arm clouds (blue squares).
The black, orange, and blue dashed lines represent the fitted results for the full sample, on-arm clouds, and inter-arm clouds, respectively.
The vertical solid line marks the lower truncation limit of $3.5 \times 10^{5} M_{\odot}$ for the full sample and on-arm clouds, while the vertical dotted line marks the lower truncation limit of $1.0 \times 10^{5} M_{\odot}$ for the inter-arm clouds.
}
\label{fig:ms}
\end{figure}

The GMC mass spectrum provides an equivalent description of how molecular gas is organized into cloud structures of varying masses. 
Variations in the mass spectrum across different regions may reflect differences in the processes that regulate the formation, evolution, and destruction of clouds \citep{Rosolowsky2005PASP,Colombo2014ApJ,Keilmann2024A&A}.
The GMC mass spectrum is typically expressed in its differential form and modeled as a power law:
\begin{equation}
f_{M} = \frac {dN}{dM} \propto M^{\gamma},
\end{equation}
The integral of this expression yields the cumulative mass distribution, which represents the number of GMCs with masses greater than a reference mass $M$ as a function of that reference mass:
\begin{equation}
N_{M>M_{0}} = (\frac {M}{M_{0}} )^{\gamma + 1},
\end{equation}
To account for deviations at the high-mass end, it is often useful to model the mass spectrum using a truncated power law \citep{Williams1997ApJ,Colombo2014ApJ}:
\begin{equation}
N_{M>M_{0}} = N_{0} [(\frac {M}{M_{0}} )^{\gamma + 1}-1],
\end{equation}
where $M_{0}$ is the maximum mass in the distribution and $N_{0}$ is the number of GMCs more massive than $2^{1/(\gamma +1)}M_{0}$ \citep{Williams1997ApJ}, the truncation mass where the distribution deviates from a simple power law.

For this study, the lower truncation limit was set at $3.5 \times 10^{5} M_{\odot}$ for the full sample and on-arm clouds, and $1.0 \times 10^{5} M_{\odot}$ for the inter-arm clouds.
These values were chosen to ensure that the cloud mass distributions follow a power-law form above the thresholds (see the top left panel of Figure \ref{fig:distribution}). 
Setting the limit lower introduces incompleteness and biases the fitted slopes, whereas adopting higher limits does not significantly change the results. 
The higher truncation limit adopted for the on-arm sample reflects the elevated background levels in these regions, which reduce the detectability of low-mass sources \citep{Kirk15}; in addition, source blending among GMCs in high-background environments further contributes to incompleteness.
The upper truncation limit was defined as the cloud mass corresponding to $N$ = 5 to minimize the impact of limited sampling at the high-mass end.
Figure \ref{fig:ms} presents the cumulative mass spectra for each group, classified by the arm mask introduced in Section \ref{sec:arm}, alongside the fitted results using the truncated power-law model.
The slope values were determined to be $\gamma = -1.84$ for the inter-arm GMCs, $\gamma = -1.80$ for the on-arm clouds, and $\gamma = -1.89$ for the full sample.
Although the $\gamma$ values are similar across the groups,  extremely massive clouds ($M >$ $10^{6} M_{\odot}$) are primarily located within the spiral arms. 
In contrast, the inter-arm regions contain only three clouds with masses exceeding $10^{6} M_{\odot}$.
This distribution results in a slightly shallower slope for the mass spectrum in the spiral arms compared to the inter-arm regions.

The power-law slopes derived in this study are consistent with previous findings for GMCs in the Milky Way, which typically range between $-1.6$ and $-2.5$ \citep[e.g.,][]{Sanders1985ApJ,Solomon1987ApJ,Roman-Duval2010ApJ,Rice2016ApJ,Miville2017ApJ,Fujita2023PASJ}. 
Our estimate of $\gamma$ aligns with the value of -1.55 obtained from high-resolution CO mapping of a subset of M31 GMCs observed with the BIMA interferometer \citep{Blitz2007}. 
It is also comparable to the value reported for GMCs in M33 \citep[][]{Gratier2012A&A,Keilmann2024A&A} and other nearby galaxies \citep[e.g., $\gamma$$\lesssim$$-2$,][]{Rosolowsky2021MNRAS}.

\subsection{Mass-Size Relation}

The GMC mass-size relation fundamentally describes the relationship between the mass and size of a population of clouds, which may be linked to the properties of interstellar turbulence \citep{Kritsuk2007ApJ}. 
The seminal work by \cite{Larson1981MNRAS} established an important observational relationship between size and volume density, expressed as $n_{\mbox{\scriptsize H}}$ $\propto$ $R_{c}$$^{-1}$.
This relationship implies that $M$ $\propto$ $R_{c}$$^{2}$, suggesting that the surface density of molecular clouds is independent of $R_{c}$.
Both numerical simulations and observations of Galactic GMCs have reported mass-size relations with similar or slightly higher slopes, typically ranging from 1.9 to 2.5 \citep[e.g.,][]{Kritsuk2007ApJ,Federrath2009ApJ,Lombardi2010A&A,Roman-Duval2010ApJ,Miville2017ApJ,Lada2020ApJ,2020MNRAS.493..351C,2022ApJ...931....9L}.

\begin{figure}
\includegraphics[width=1.0\linewidth]{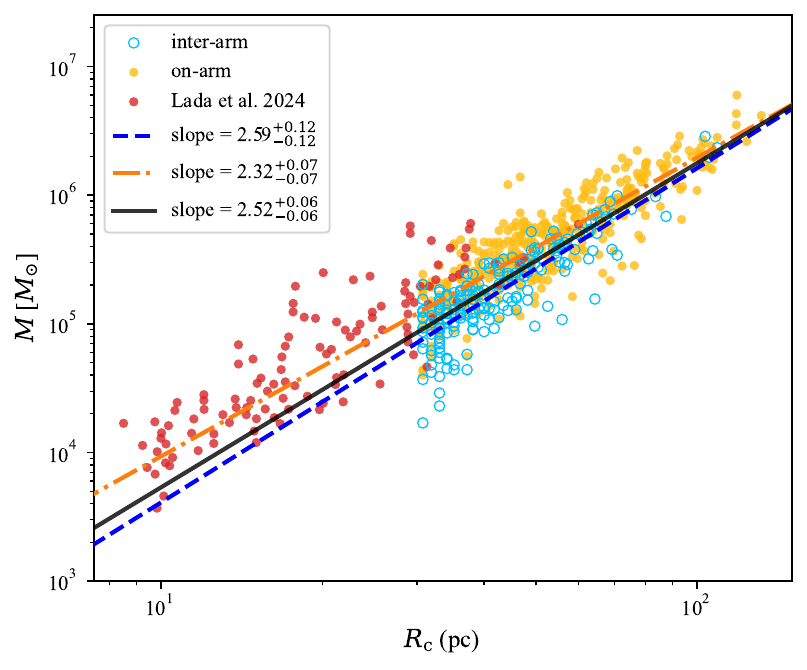}
\caption{
Mass-size relation of the identified GMCs.
The black solid line, orange dashed-dotted line, and blue dashed line represent the fitted results for the full sample, on-arm clouds, and inter-arm clouds, respectively.
The red points are the measurements derived from $^{12}$CO observations with the SMA \citep{Lada2024ApJ}; their systematically smaller sizes reflect the higher angular resolution of those data compared to this work.
}
\label{fig:m_vs_r}
\end{figure}

Figure \ref{fig:m_vs_r} presents the mass-size relation for our sample of molecular clouds. 
A clear and well-defined mass-size scaling relation is observed for the M31 clouds in our samples. 
Performing a linear regression for the full sample yields the following correlation (represented by the black dashed line): 
\begin{equation} \label{eq:mass_size1}
\log(M/M_{\odot}) = 2.52^{+0.06}_{-0.06} \log(R_{\rm c}/{\rm pc}) + 1.21,
\end{equation} 
with a Spearman coefficient of 0.87. 
The mass-size scaling relation persists for both inter-arm clouds and on-arm clouds.
For on-arm clouds, a least-squares linear fit gives:
\begin{equation} \label{eq:mass_size2}
\log(M/M_{\odot}) = 2.32^{+0.07}_{-0.07} \log(R_{\rm c}/{\rm pc}) + 1.65.
\end{equation}
For inter-arm clouds, the relation is:
\begin{equation} \label{eq:mass_size3}
\log(M/M_{\odot}) = 2.59^{+0.09}_{-0.09} \log(R_{\rm c}/{\rm pc}) + 1.01.
\end{equation}
The measured slope values are comparable to those reported for clouds observed by the SMA in M31 \citep{Lada2024ApJ}. 
We note that the GMCs identified by \citet{Lada2024ApJ} are systematically smaller than those in our catalog, which is primarily a consequence of the higher data resolution used in their study.

\begin{figure*}
\begin{tabular}{c}
\hspace{-0.5cm}\includegraphics[width=17.45cm]{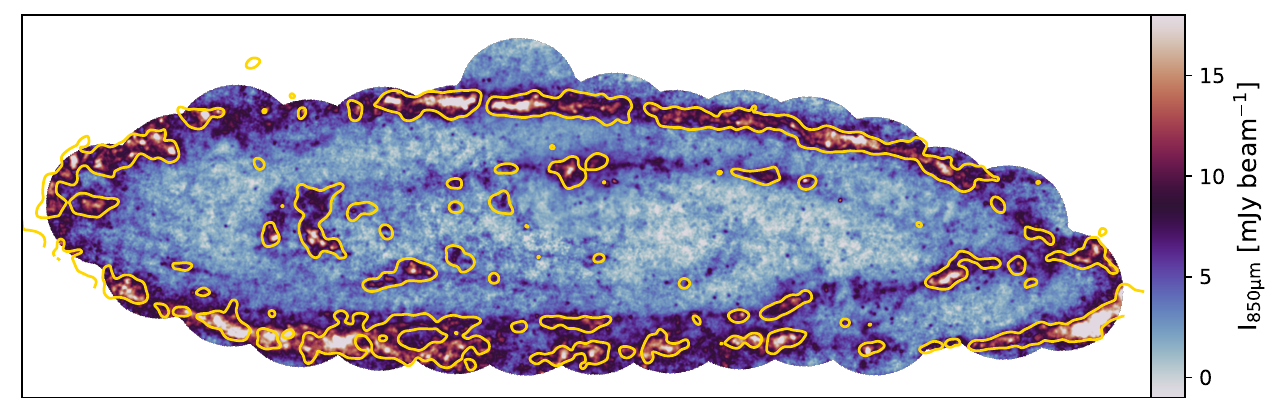} \\
\end{tabular}
\begin{tabular}{c}
\hspace{0.25cm}\includegraphics[width=18.cm]{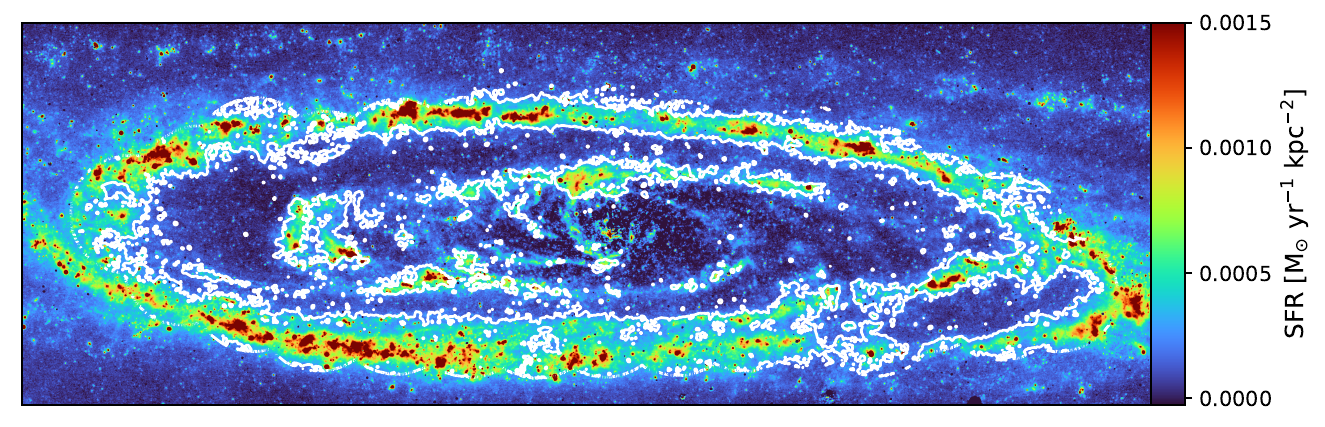} \\
\end{tabular}
\caption{
Comparison of the JCMT-SCUBA2 850 $\mu$m dust emission (color map) and the IRAM 30m CO intensity map (orange contours) and comparison between the SFR map from \cite{ford2013ApJ} (color map) and the JCMT-SCUBA2 850 $\mu$m dust emission (white contours).
}
\label{fig:co_and_sfr}
\end{figure*}

Although the GMC population in our sample exhibits a clear mass-size relation, the measured slope of the relation deviates significantly from 2 with an appreciable scatter.
The scatter is likely due to variations in the surface densities of the GMCs, which are not uniform (see Section \ref{sec:distribution}).
The $\Sigma_{\mbox{\scriptsize dust}}$ distributions for on-arm clouds, inter-arm clouds, and the full sample all display single-peaked distributions, primarily ranging from 0.3 to 1.0 $M_\odot$ pc$^{-2}$.
Additionally, the on-arm clouds are systematically denser than the inter-arm clouds (see Figure \ref{fig:distribution} and the scatter in Figure \ref{fig:m_vs_r}).

\subsection{Comparison with Ancillary Observations}

\subsubsection{CO Observations}\label{subsubsec:co}

Besides dust thermal emissions, low-$J$ CO rotational line observations are another common way of probing GMCs, and there have been rich CO detections towards M31, both single dish telescope observations, e.g., IRAM 30m \citep[][]{Nieten2006A&A} and JCMT \citep{Li2020MNRAS,SmithApJS}, and interferometer observations, e.g., BIMA \citep[][]{Rosolowsky2007ApJ,Sheth2008ApJ}, CARMA \citep{Caldu-Primo2016AJ}, and SMA \citep{Forbrich2020ApJ}.
However, because of the large angular scale of M31 on the sky, most observations mainly focus on the bright spiral arm regions.
The only subarcminute resolution CO survey across the entire disk of M31 is made by \cite{Nieten2006A&A} using IRAM 30m.

This 30m $^{12}$CO 1-0 map is presented in the top panel of Figure \ref{fig:co_and_sfr}.
The top panel of Figure \ref{fig:co_and_sfr} compares the JCMT-SCUBA2 850 $\mu$m dust emission (color scale) and the IRAM 30m CO intensity map (5$\sigma$ contours in orange).
While the CO survey clearly detects the prominent 5 kpc and 10 kpc rings, only a small fraction of inter-arm GMCs show CO counterparts.
In the bright arm regions, we observe a typical 850 $\mu$m-to-CO ratio of $\sim$ 1 mJy beam$^{-1}$ / K km s$^{-1}$.
Assuming a similar ratio applies to the inter-arm GMCs, with peak 850 $\mu$m flux densities higher than 3 mJy beam$^{-1}$, their expected CO intensities would be around 3 K km s$^{-1}$, corresponding to $\sim$8$\sigma$ detections. 
However, the majority of inter-arm clouds remain undetected in this CO survey.
This discrepancy may be due to a lower molecular gas fraction in the inter-arm regions compared to the spiral arms (as previously reported in the Milky Way \citep[e.g.,][]{wang2020A&A...634A..83W}) and/or a variation in the dust-to-gas ratio between on-arm and inter-arm environments. 
Additionally, the relatively coarse angular resolution (23$''$, compared to 14$''$ for the JCMT-SCUBA-2 850 $\mu$m map) of the CO data could dilute the CO emissions from the compact inter-arm clouds.
A deeper CO survey specifically targeting the inter-arm regions is necessary for a more detailed comparison between on-arm and inter-arm GMCs.

\subsubsection{Star Formation Rate}

In this subsection, we investigate the relationship between the SFR ($\mathrm{\Sigma_{SFR}}$) and gas surface density ($\mathrm{\Sigma_{gas}}$), commonly referred to as the ``star-formation law" \citep{Kennicutt1998}, assuming the following form:
\begin{equation} \label{eq:KS}
\mathrm{\Sigma_{SFR}} = A \mathrm{\Sigma_{gas}^{N}},
\end{equation} 
where $N$ is the power-law index and $A$ is a normalization factor related to the star formation efficiency.
We computed $\mathrm{\Sigma_{gas}}$ for each cloud from $\Sigma_{\mathrm{dust}}$ using the dust-to-gas ratio from \citet{Draine2014ApJ}, following Equation~\ref{eq:N_gas}.
Most of the clouds have $\Sigma_{\mathrm{gas}}$ values in the range 10 to 100 $M_\odot$ pc$^{-2}$, which lies above the surface density threshold where atomic gas saturates \citep[e.g., $\sim 5-10\ M_\odot,\mathrm{pc}^{-2}$;][]{Bigiel2008AJ,Krumholz2008ApJ...689..865K,Krumholz2009ApJ...693..216K,Sternberg2014ApJ...790...10S}. 
Therefore, the majority of these clouds are expected to be dominated by molecular gas.
Using the SFR distribution map (see bottom panel of Figure \ref{fig:co_and_sfr}) of M31 derived from 24 $\mu$m observations with the {\it Spitzer Space Telescope} \citep{Gordon2006ApJ} and FUV observations from {\it GALEX} \citep{Thilker2005ApJ} at an angular resolution of 6$''$ \citep{ford2013ApJ}, we derived mean value within the cloud as the SFR surface density ($\mathrm{\Sigma_{SFR}}$) for each GMC and examine its relationship with gas surface density ($\mathrm{\Sigma_{gas}}$).
Before comparing $\mathrm{\Sigma_{SFR}}$ and $\mathrm{\Sigma_{gas}}$, the SFR distribution map was convolved to an angular resolution of 14$''$ (e.g., the angular resolution of the JCMT-SCUBA2 850 $\mu$m map), with a typical 3$\sigma$ uncertainty of 1.5 $\times$ 10$^{-4}$ $M_\odot$ yr$^{-1}$ kpc$^{-2}$ for the target region.

\begin{figure}
\includegraphics[width=1.0\linewidth]{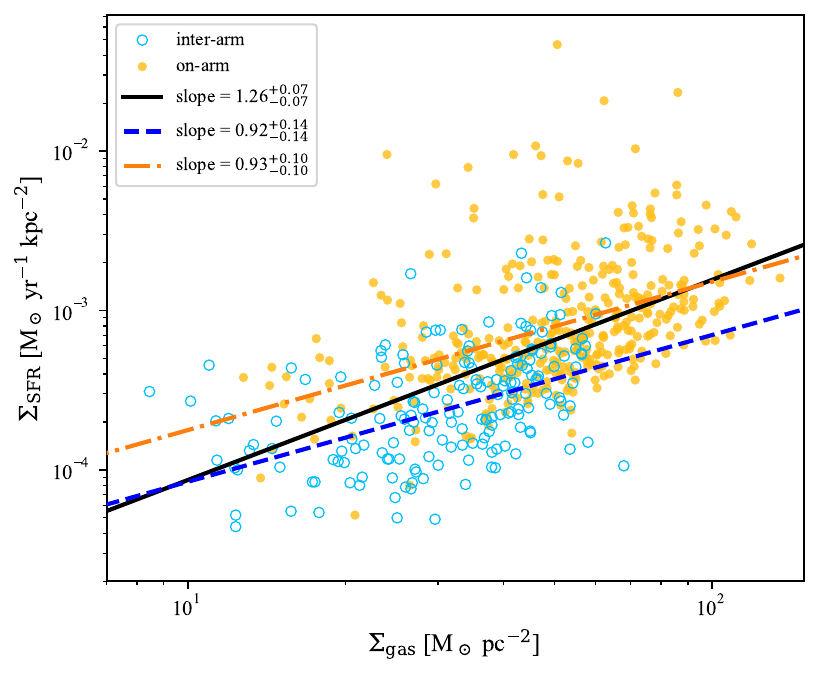}
\caption{
Star formation rate surface density against gas surface density. 
The black solid line, orange dashed-dotted line, and blue dashed line represent the fitted results for the full sample, on-arm clouds, and inter-arm clouds, respectively.
}
\label{fig:KS}
\end{figure}

Figure \ref{fig:KS} shows the SFR versus gas surface density for our sample of molecular clouds.
We performed a linear fit in order to find the index, $N$, from Equation \ref{eq:KS}.
For the full sample, we found superlinear relationships ($N > 1$), with $N \sim 1.26$ for the full sample, $N \sim 0.93$ for the on-arm clouds, and $N \sim 0.93$ for the inter-arm clouds.
However, several caveats should be noted when interpreting the index $N$ in this work.
First, the slope fitting is highly sensitive to the particular methodology used and limited by the sensitivity of $\mathrm{\Sigma_{SFR}}$ \citep[e.g.,][]{Leroy2013AJ....146...19L}, particularly for the inter-arm clouds.
Second, the dynamic range of $\mathrm{\Sigma_{gas}}$ is relatively narrow, spanning primarily 10 to 100 $M_\odot$ pc$^{-2}$.
Moreover, previous works have shown that the $\mathrm{\Sigma_{SFR}}$-$\mathrm{\Sigma_{gas}}$ relation varies with galactocentric radius in M31 \citep[e.g.,][]{ford2013ApJ,Rahmani2016MNRAS.456.4128R}, which may introduce additional complexity into slope fitting for both the on-arm and inter-arm cloud samples.

A systematically higher $\Sigma_{\rm SFR}$ (by $\sim$0.8 dex) is observed in the on-arm regions compared to the inter-arm regions at a given $\Sigma_{\rm gas}$, particularly within the range of $\Sigma_{\rm gas} = 20$–$50\ M_\odot\ \mathrm{pc}^{-2}$.
The lower $\Sigma_{\rm SFR}$ for the inter-arm clouds may reflect suppressed star formation activity, consistent with the lower dust contents and temperatures observed in inter-arm clouds. 
In contrast, the elevated $\Sigma_{\rm SFR}$ in the on-arm clouds of M31 likely results from shock compression induced by the spiral arms \citep{Roberts1969}, as predicted by density wave theory \citep{Lin1964}.
Such spiral shocks could produce denser GMCs, which are observable on the arms of M31. 
\cite{Yu2021} further showed that spiral arms could increase the global star formation efficiency in the cold gas reservoir (see also \cite{Kendall2015} and \cite{Yu2022}), which in turn maintains the spiral arms. 
However, the role of spiral arms in triggering star formation remains a topic of active debate.
While some studies report enhanced star formation activity along spiral arms \citep[e.g.,][]{Lord1990ApJ...356..135L,Ragan2018MNRAS.479.2361R,Yu2021,Yu2022,Chen2024MNRAS.534..883C}, other works do not find a significant difference with the inter-arm regions \citep[e.g.,][]{Foyle2010ApJ...725..534F,Querejeta2021A&A...656A.133Q,Sun2024ApJ...973..137S,Romanelli2025A&A...698A.296R}.
Further observational and theoretical studies are needed to better understand the impact of spiral arms on star formation.


\begin{figure*}
\begin{tabular}{c c}
\hspace{-0.6cm}\includegraphics[width=0.49\linewidth]{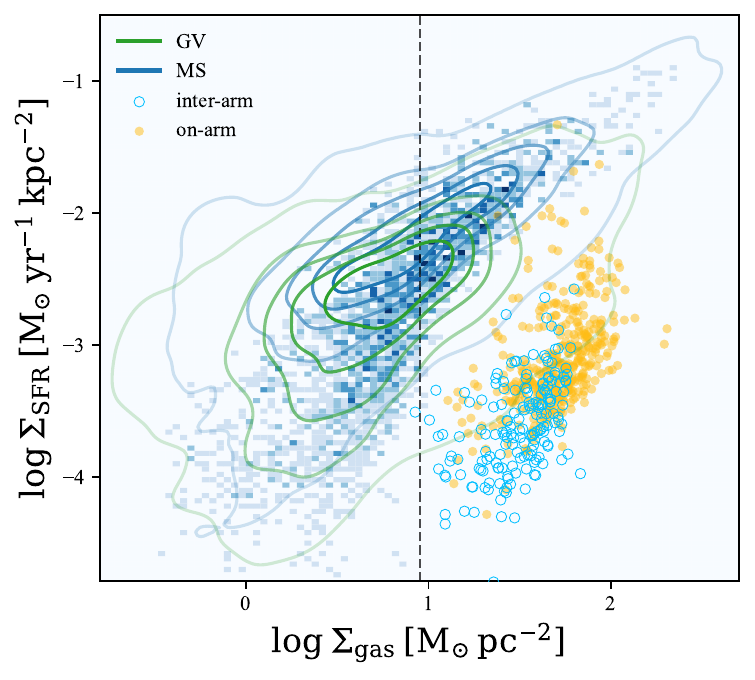}
\includegraphics[width=0.49\linewidth]{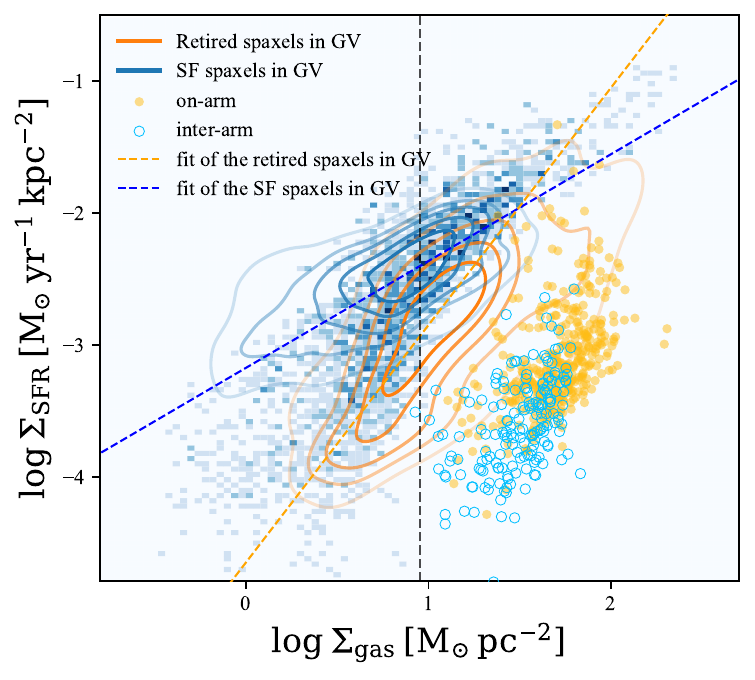}
\end{tabular}
\begin{tabular}{c c}
\includegraphics[width=0.49\linewidth]{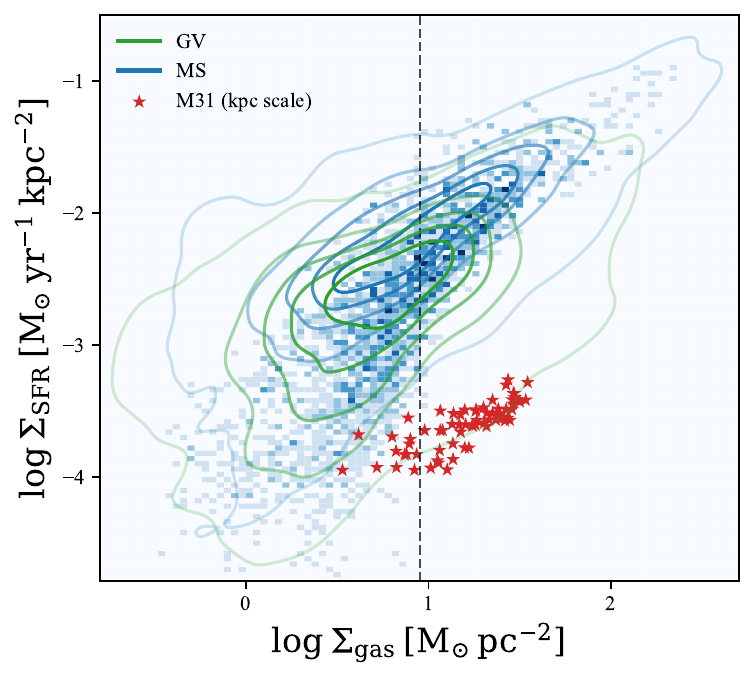}
\includegraphics[width=0.49\linewidth]{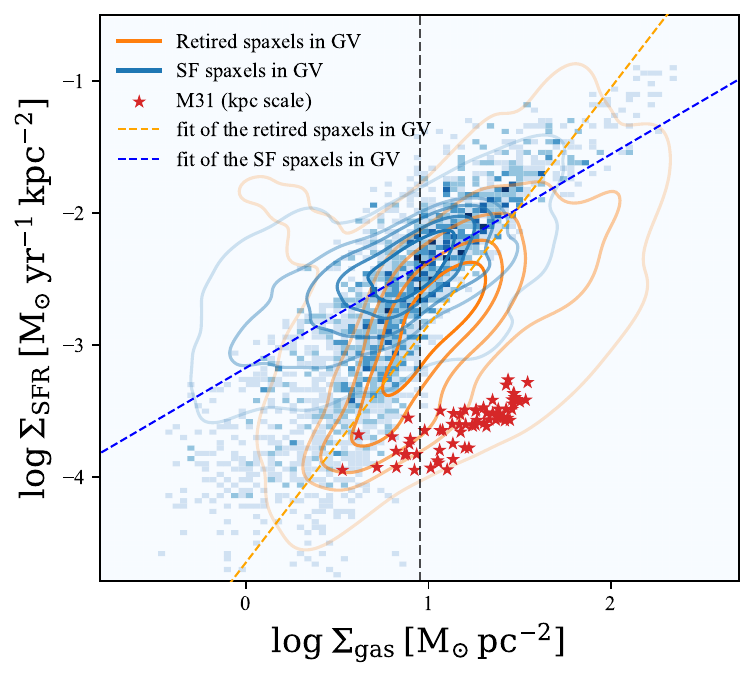}
\end{tabular}
\caption{
(Top panels) The Kennicutt–Schmidt relation.
Left: The background blue histogram shows the number densities of nearby galaxy measurements at sub-kpc scales from \cite{Bigiel2008AJ}.
Blue and green contours mark kpc-scale measurements from the ALMaQUEST survey for MS and GV galaxies \citep{Lin2022ApJ...926..175L}.
Orange and blue points represent M31 GMCs measurements.
Contours enclose 2\%, 20\%, 40\%, 60\%, and 80\% of the density peaks. 
Dashed vertical line shows saturation surface density at 9 $M_\odot$ pc$^{-2}$ \citep{Bigiel2008AJ}.
Right: Same as left, but with blue contours for SF spaxels and orange contours for retired spaxels in GV galaxies.
(Bottom panels) Same as top, but for kpc-scale measurements in M31.
\label{fig:msgv}}
\end{figure*}

The SFR of M31 is uniquely low compared to other Local Group spiral galaxies, even within its spiral arms \citep{Pagani1999A&A}.
To address how effectively gas forms stars using GMC-scale measurements, we calculate the gas depletion time defined as $t_{\mbox{\scriptsize dep}} \equiv \Sigma_{\mbox{\scriptsize gas}} / \Sigma_{\mbox{\scriptsize SFR}}$.
Most depletion times fall in the range from 20 to 100 Gyr, with a median value of 87 Gyr.
The median value is over an order of magnitude longer than the typical values observed in other nearby galaxies \citep[e.g.,][]{Leroy2008AJ,Sun2023ApJ...945L..19S}.
The low star formation efficiency (SFE, defined as SFE = $1/t_{\mbox{\scriptsize dep}}$) of M31 has been noted in previous studies \citep[e.g.,][]{Mutch2011ApJ,Davidge2012ApJ,Kelvin2018MNRAS}, and the galaxy is commonly classified as residing in the so-called green valley (GV).
Galaxies in the GV occupy an intermediate region in the color–magnitude or SFR–stellar mass diagram and are generally thought to be transitioning from star-forming to passive, in a process referred to as quenching \citep{Salim2007ApJS,Wyder2007ApJS,Lin2017ApJ...851...18L}. 

To further explore this, we compare the GMC-scale measurements in M31 with kpc-scale measurements from the ALMA-MaNGA QUEnch and STar formation (ALMaQUEST) survey \citep{Lin2019ApJ...884L..33L}, as well as sub-kpc-scale measurements in nearby galaxies from \citet{Bigiel2008AJ}, in the $\Sigma_{\mbox{\scriptsize gas}}$–$\Sigma_{\mbox{\scriptsize SFR}}$ plane (see top panels of Figure~\ref{fig:msgv}).
To ensure consistency, we multiplied the SFR values from the ALMaQUEST survey by a factor of 0.61 to match the IMF assumption used in both \citet{Bigiel2008AJ} and \citet{ford2013ApJ}.
The M31 GMCs clearly lie below the distribution of the MS galaxy and are primarily located in the lower-right region of the GV galaxy distribution, indicating lower SFEs.
The left panel further separates GV galaxies into star-forming (SF; blue contours) and retired (orange contours) spaxels.
Remarkably, most M31 GMCs fall within the lower-right portion of the retired spaxel distribution.
Moreover, despite a systematic offset, the M31 GMCs and retired GV spaxels exhibit a similar shape and slope in the $\Sigma_{\mbox{\scriptsize gas}}$–$\Sigma_{\mbox{\scriptsize SFR}}$ plane. 
This suggests that star formation in M31 GMCs behaves similarly to that in the retired regions of GV galaxies and may have already ceased or is significantly suppressed.

One plausible explanation for why star formation efficiency is so slow for the GMCs in M31 is that, while GMCs in M31 contain substantial amounts of gas, only the gravitationally bound component is actively collapsing to form stars.
Star formation in the GMCs is a complex process influenced by many physical interactions \citep[e.g.,][]{Shu1991ApJ,McKee2007ARA&A}.
Observationally, key factors such as gravity, turbulence, magnetic fields, and star-formation feedback are deeply intertwined, complicating the accurate identification of the fraction of gas actively involved in star formation.
By analyzing the column density probability distribution function (N-PDF), \citet{Jiao2025arXiv} isolated the gravity-dominated gas component in a sample of star-forming molecular clouds in the Milky Way.
\citet{Jiao2025arXiv} found that gravitationally bound gas plays a decisive role in star formation: the mass of bound gas sets the SFR, and this component appears to have a nearly universal star formation efficiency.
This hypothesis naturally explains the very low SFE in M31’s GMCs if these GMCs are highly turbulent and have a small fraction of gravitationally bound gas.
Additional support for this scenario comes from recent spatially resolved $^{12}$CO and $^{13}$CO observations of GMCs in M31, which reveal that a large fraction (57\%) of these clouds appear to be gravitationally unbound \citep{Lada2024ApJ, Lada2025ApJ}.
This finding is consistent with results from the Milky Way, where \citet{Evans2021ApJ} suggest that as many as $\sim$80\% of GMCs may also be unbound. 

Follow-up investigations of GMCs in M31 could offer valuable insights into the physical mechanisms responsible for quenching star formation at sub-galactic scales.
By examining the dynamical states, internal structures, and feedback processes in detail within these clouds, such studies may reveal how star formation is suppressed in otherwise gas-rich environments, thereby advancing our understanding of the broader processes that regulate galaxy evolution.

The SFR estimates used in this work are adopted from \cite{ford2013ApJ}, which are derived based on {\it Spitzer} 24 $\mu$m and {\it GALEX} FUV observations.
We caution that using this FUV+IR hybrid method to calculate SFRs on GMC scales may be problematic, due to stochastic IMF sampling, non-uniform star formation histories, and contamination from diffuse emission.
These limitations have been well documented in the literature \citep[e.g.,][]{Kennicutt2012ARA&A..50..531K,Leroy2013AJ....146...19L,Kruijssen2014MNRAS.439.3239K,Schinnerer2024ARA&A..62..369S} and may lead to artificially long depletion times in GMC-scale analyses.
To mitigate these issues, we also repeated the analysis at 1 kpc resolution by convolving both the gas and SFR surface density maps. 
At this 1 kpc scale, the derived SFRs are more robust and directly comparable to kpc-scale studies in the literature. 
The resulting M31 data points at 1 kpc spatial resolution (red stars in bottom panels of Figure~\ref{fig:msgv}) remain offset below the distribution of main-sequence galaxies and are primarily located in the lower-right region of the green valley distribution.
Relative to the literature, the gas surface densities we derived (primarily ranging from 10$^{0.5}$ to 10$^{1.5}$ $M_\odot$ pc$^{-2}$) are generally consistent with the values reported in \cite{Braun2009ApJ...695..937B}.
Remaining differences with other studies are likely due to methodological choices and modeling assumptions used to derive the gas surface density.
For example, \cite{Braun2009ApJ...695..937B} reported molecular gas surface densities that are a factor of 1.6 higher than the values of \cite{Tabatabaei2010A&A...517A..77T} due to differences in scaling of the CO data, inclination, angular resolution, and radial range; moreover, the dust surface densities estimated using the same dust model in \cite{Draine2014ApJ} may be systematically high by a factor of 2 \citep{Rahmani2016MNRAS.456.4128R}.

For the SFR surface densities, our 1 kpc-scale measurements are about 0.5 dex lower than those in \cite{Braun2009ApJ...695..937B} and \cite{Tabatabaei2010A&A...517A..77T}.
This offset is plausibly attributable to the choice of SFR tracers:  \cite{Braun2009ApJ...695..937B} used total infrared and FUV, \cite{Tabatabaei2010A&A...517A..77T} used H${\alpha}$, whereas we adopt FUV+24$\mu$m \citep{ford2013ApJ}.
A systematic offset of similar magnitude between SFR surface densities derived from different tracers in M31 is reported by \citet{Rahmani2016MNRAS.456.4128R}.
In addition, \citet{ford2013ApJ} applied a correction to remove foreground and evolved-stellar emission when estimating the SFR, which reduces the measured global SFR by $\sim$25\%.
We note that even after applying a +0.5 dex shift to the SFR surface densities, our kpc-scale measurements in M31 show systematically lower star formation efficiencies compared to those reported for nearby galaxies on similar scales by \cite{Bigiel2008AJ}. 
Such a difference is qualitatively consistent with M31’s classification as a green valley galaxy \citep[e.g.,][]{Kelvin2018MNRAS}.


\section{Summary}\label{section:summary}

In this paper, we present new JCMT-SCUBA2 450/850 $\mu$m observations of M31.
Combined with {\it Herschel} and {\it Planck} data, we analyzed the dust and gas properties in M31 and identified a total of 572 GMCs using the {\it Dendrogram} algorithm. 
Our main findings are as follows:

\begin{enumerate}
\item Within the field of view of our new JCMT-SCUBA2 observations, we find that the dust mass surface density in M31 peaks in two prominent rings, at R $\sim$ 5 kpc and R$\sim$10 kpc, consistent with previous observations.
The total dust mass is estimated to be 2.7$\times$10$^{7}$ $M_{\odot}$, while the inter-arm dust mass within the field of view is 1.0$\times$10$^7$ $M_{\odot}$ (accounting for 37\% of the total dust mass in our mapped region).  
Using the total dust mass of 5.4$\times$10$^7$ $M_{\odot}$ for M31 reported by \cite{Draine2014ApJ}, we find that approximately 19\% of M31's dust resides in the inter-arm region within the 10 kpc ring.
\item Consistent with previous works, the dust temperature in the central region (R $\lesssim$ 3 kpc) is systematically higher than in the outer region (R $\gtrsim$ 3 kpc).
A noticeable decrease in dust temperature is observed, dropping from $\sim$ 30 K to $\sim$ 17 K, from the center to the $\sim$ 3 kpc radii.
Beyond $\sim$ 3 kpc, the dust temperature remains relatively stable at around 17 K.
\item The inclusion of longer wavelength data in the SED fitting yields a median $\beta$ value of $\sim$ 1.8, with smaller variations in $\beta$ (1.5 $\lesssim$ $\beta$ $\lesssim$ 2.3), particularly for R $\lesssim$ 7 kpc.
We observe a relatively uniform distribution of $\beta$ and no significant correlation between $\beta$ and radius.
\item For the full sample of GMCs identified in this work, the median values are as follows: median mass is 3.0$\times$10$^5$ $M_{\odot}$, median dust temperature is 17.5\,K, median $\beta$ is 1.8, median radius is  46.7\,pc, median dust surface density is 0.45 $M_\odot$\,pc$^{-2}$, and median gas mean volume density is 19.4 cm$^{-3}$.
These values are comparable to measurements observed in the Milky Way.
\item To investigate the environmental dependence of GMC properties in M31, we divided the JCMT field of view into two regions: the arm environment and the inter-arm environment.
This resulted in a classification of 383 on-arm GMCs and 189 inter-arm GMCs.
Our analysis reveals a clear dependence of GMC properties on their galactic environments.
Inter-arm clouds are systematically less massive, more diffuse, colder, and more quiescent compared to on-arm GMCs in M31.
\item We fit the power-law slope of the cumulative mass function for the clouds.
The slope values for the inter-arm clouds, on-arm clouds, and the full sample are similar ($\gamma$ = -1.84 for the inter-arm GMCs, $\gamma$ = -1.80 for the on-arm clouds, and $\gamma$ = -1.89 for the full sample).
However, the inter-arm region lacks the more massive clouds ($M >$ $5 \times 10^{5} M_{\odot}$).
\item A strong correlation is observed between the masses and sizes of GMCs in M31, following $M$ $\sim$ $R_{c}$$^{2.52}$.
This result aligns with recent observations on GMCs in both the Milky Way and other nearby galaxies but differs from the seminal findings of \cite{Larson1981MNRAS}.
\item We investigate the relationship between $\Sigma_{\mathrm{SFR}}$ and $\Sigma_{\mathrm{gas}}$ for the identified GMCs and find a superlinear trend, with a power-law index of $N \sim 1.22$ for the full sample.
The star formation efficiency at the GMC scale in M31 is significantly lower than that observed in main-sequence galaxies and is comparable to the low-SFE end of green valley galaxies.
By repeating the analysis at 1 kpc spatial resolution, we find that M31 remains offset below the star-forming main sequence with systematically low SFEs, but with reduced scatter compared to the GMC-scale measurements.
\end{enumerate}

\begin{acknowledgments}
The James Clerk Maxwell Telescope is operated by the East Asian Observatory on behalf of The National Astronomical Observatory of Japan; Academia Sinica Institute of Astronomy and Astrophysics; the Korea Astronomy and Space Science Institute; the National Astronomical Research Institute of Thailand; Center for Astronomical Mega-Science (as well as the National Key R\&D Program of China with No. 2017YFA0402700). 
Additional funding support is provided by the Science and Technology Facilities Council of the United Kingdom and participating universities and organizations in the United Kingdom and Canada.
Additional funds for the construction of SCUBA-2 were provided by the Canada Foundation for Innovation.
{\it Herschel} is an ESA space observatory with science instruments provided by European-led Principal Investigator consortia and with important participation from NASA.

S.J. is supported by NSFC grant nos. 12588202 and 12041302, by the National Key R\&D Program of China No. 2023YFA1608004.
H.B.L. is supported by the National Science and Technology Council (NSTC) of Taiwan (Grant Nos. 111-2112-M-110-022-MY3, 113-2112-M-110-022-MY3).
Y.C. was partially supported by a Grant-in-Aid for Scientific Research (KAKENHI  number JP24K17103) of the JSPS. 
J.d.B. acknowledges support from the Smithsonian Institution as Submillimeter Array (SMA) Fellows.
Z.Z. is also supported by  2023YFC2206403, 2024YFA1611602; NSFC 12373012, 12041302; and CMS-CSST-2025-A08.
N.K.Y. is supported by the project funded by China Postdoctoral Science Foundation No. 2022M723175 and GZB20230766.
\end{acknowledgments}

\facility{Herschel, Planck, JCMT}
\software{CASA, Numpy, APLpy}

\appendix

\section{Example of Cloud Identification}\label{gmc_example}

\begin{figure*}
\vspace{-0.1cm}
\begin{center}
\includegraphics[width=17cm]{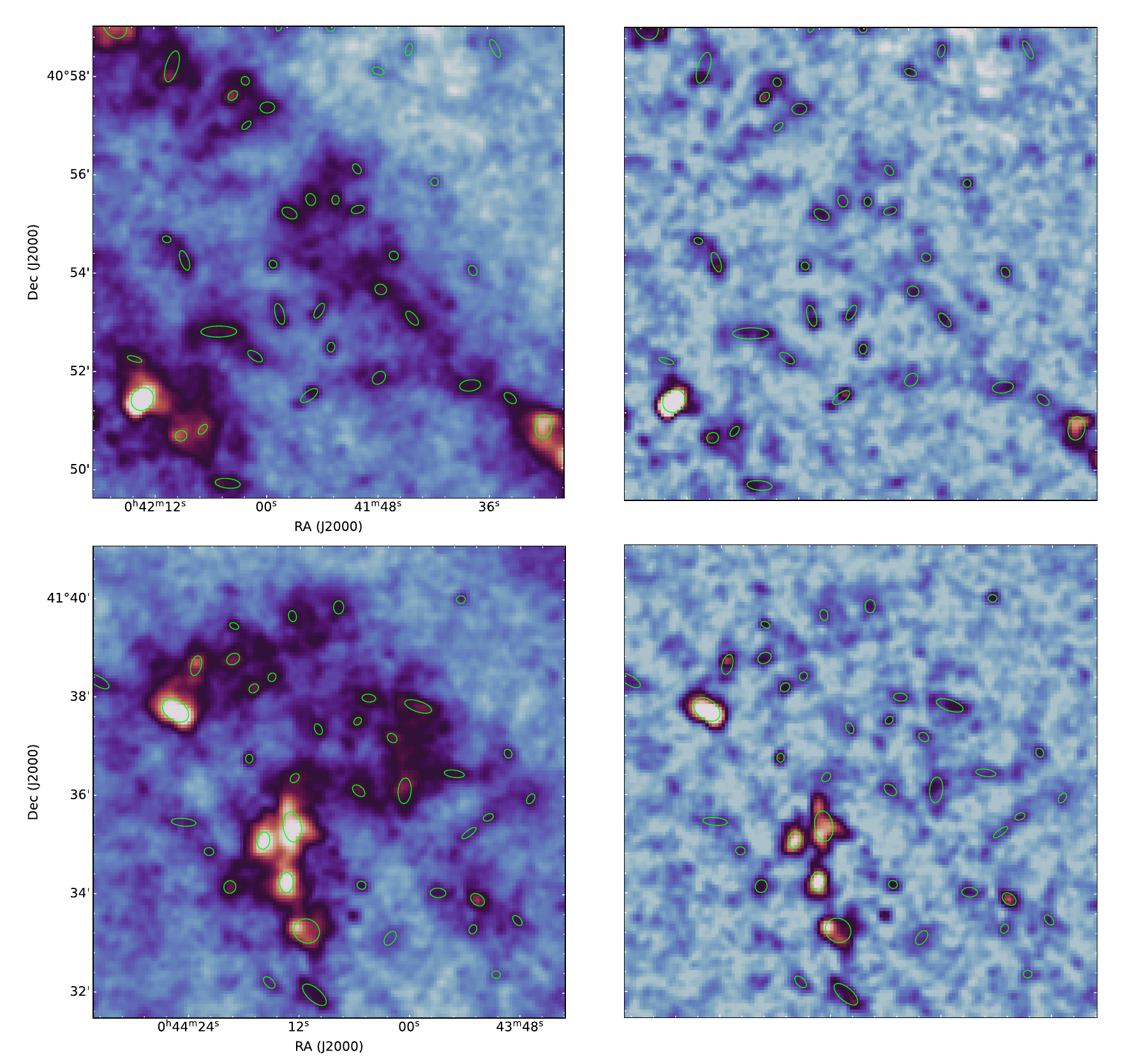}
\end{center}
\caption{
Identified GMCs (lime ellipses) in two representative subregions.
The left panels show the 850 $\mu$m continuum image, while the right panels show the decomposed image.
}
\label{fig:gmc_example}
\end{figure*}

As described in Section \ref{sec:gmcs}, our GMC catalog was generated by applying the {\it Dendrogram} algorithm to the decomposed images shown in the right panels of Figure \ref{fig:gmc_example}.
We adopted the minimum flux density threshold to be 4$\sigma$ noise level of the decomposed image, the minimum significance for structures to 1$\sigma$, and the minimum area to the size of the beam (14\arcsec) to identify the leaf structure as an individual GMC.
Figure \ref{fig:gmc_example}  illustrates our source identification results for two representative subregions.
The left panels of Figure \ref{fig:gmc_example} display the original 850 $\mu$m continuum image, while the right panels present the decomposed images processed using the CDD technique \citep{Li2022ApJS}, which preserves only structures smaller than $\sim$300 pc.
Each identified GMC is represented by an ellipse showing the deconvolved major and minor axes and position angle derived from our measurements.

To ensure catalog reliability, we employed relatively conservative selection parameters.
As a result, some faint sources that might be visually identifiable may not have been confirmed as GMCs.
A more complete but potentially less reliable sample could be obtained by relaxing the input parameters.

\bibliographystyle{yahapj}
\bibliography{references}
\end{document}